\documentclass[prx,aps,superscriptaddress,footinbib,twocolumn]{revtex4-2}

\usepackage{graphicx}
\usepackage{array}
\usepackage{xcolor}
\renewcommand{\thetable}{\arabic{table}}
\usepackage{booktabs}
\usepackage{amsmath}
\usepackage{xcolor}
\usepackage{colortbl}
\usepackage[normalem]{ulem}
\newlength{\cellw}
\usepackage{booktabs}
\definecolor{myblue1}{rgb}{0.89,0.95,1}
\definecolor{myblue2}{rgb}{0.78,0.91,1}
\definecolor{myblue3}{rgb}{0.67,0.87,1}
\definecolor{myblue4}{rgb}{0.56,0.83,1}
\definecolor{myblue5}{rgb}{0.45,0.79,1}
\definecolor{myblue6}{rgb}{0.34,0.70,0.95}
\definecolor{myblue7}{rgb}{0.23,0.61,0.90}
\definecolor{myblue8}{rgb}{0.14,0.47,0.82}
\definecolor{myblue9}{rgb}{0.05,0.33,0.74}

\definecolor{myorange1}{rgb}{1.00,0.96,0.87}
\definecolor{myorange2}{rgb}{1.00,0.92,0.74}
\definecolor{myorange3}{rgb}{1.00,0.86,0.57}
\definecolor{myorange4}{rgb}{1.00,0.80,0.40}
\definecolor{myorange5}{rgb}{1.00,0.72,0.23}
\definecolor{myorange6}{rgb}{1.00,0.63,0.00}
\definecolor{myorange7}{rgb}{0.94,0.53,0.00}
\definecolor{myorange8}{rgb}{0.87,0.40,0.00}
\definecolor{myorange9}{rgb}{0.75,0.27,0.00}

\newcommand{\bluecell}[1]{%
  \ifnum#1=2 \cellcolor{myblue1}\else%
  \ifnum#1=3 \cellcolor{myblue2}\else%
  \ifnum#1=4 \cellcolor{myblue3}\else%
  \ifnum#1=5 \cellcolor{myblue4}\else%
  \ifnum#1=6 \cellcolor{myblue5}\else%
  \ifnum#1=7 \cellcolor{myblue6}\else%
  \ifnum#1=8 \cellcolor{myblue7}\else%
  \ifnum#1=9 \cellcolor{myblue8}\else%
  \ifnum#1=10 \cellcolor{myblue9}\else%
  \cellcolor{white}\fi\fi\fi\fi\fi\fi\fi\fi\fi%
  \makebox[\cellw][c]{\strut #1}%
}

\newcommand{\orangecell}[1]{%
  \ifnum#1=2 \cellcolor{myorange1}\else%
  \ifnum#1=3 \cellcolor{myorange2}\else%
  \ifnum#1=4 \cellcolor{myorange3}\else%
  \ifnum#1=5 \cellcolor{myorange4}\else%
  \ifnum#1=6 \cellcolor{myorange5}\else%
  \ifnum#1=7 \cellcolor{myorange6}\else%
  \ifnum#1=8 \cellcolor{myorange7}\else%
  \ifnum#1=9 \cellcolor{myorange8}\else%
  \ifnum#1=10 \cellcolor{myorange9}\else%
  \cellcolor{white}\fi\fi\fi\fi\fi\fi\fi\fi\fi%
  \makebox[\cellw][c]{\strut #1}%
}

\usepackage{amsthm,amsmath,amssymb}
\usepackage{mathrsfs}

\newcommand{\ket}[1]{\vert{ #1 }\rangle}

\usepackage{amsthm}
\newtheorem{theorem}{Theorem}

\theoremstyle{definition}
\newtheorem{definition}{Definition}
\theoremstyle{remark}

\usepackage{amssymb}
\usepackage{graphicx}
\usepackage{epstopdf}
\usepackage{algorithm}
\usepackage[noend]{algpseudocode}
\usepackage{lipsum}

\usepackage{booktabs}
\usepackage{multirow}
\usepackage{array}
\newcolumntype{P}{>{\centering\arraybackslash}w{c}{16pt}}

\usepackage{natbib}

\usepackage[colorlinks=true,citecolor=blue,linkcolor=magenta]{hyperref}

\usepackage{xcolor}

\newcommand{\RED}[1]{{#1}}

\usepackage{comment}

\begin{document}

%\title{Practical fault-tolerant operations on high-encoding-rate planar quantum codes}
%\title{Planar Fault-Tolerant Quantum Computation with Low Overhead}
\title{Planar Fault-tolerant Logical Measurements with Low Qubit Overhead}

\author{Yingli Yang}
\affiliation{Graduate School of China Academy of Engineering Physics, Beijing 100193, China}

\author{Guo Zhang}
\affiliation{Graduate School of China Academy of Engineering Physics, Beijing 100193, China}

\author{Ying Li}
\email{yli@gscaep.ac.cn}
\affiliation{Graduate School of China Academy of Engineering Physics, Beijing 100193, China}

\begin{abstract}
\noindent\textbf{Abstract.}
Fault-tolerant quantum computation critically depends on architectures uniting high encoding rates with physical implementability. Quantum low-density parity-check (qLDPC) codes, including bivariate bicycle (BB) codes, achieve dramatic reductions in qubit overhead, yet their logical operations remain a key challenge under planar hardware constraints. Here, we introduce code craft, a framework for designing fault-tolerant logical operations on planar BB codes within a translationally invariant, two-dimensional qubit lattice. By systematically deforming codes through local modifications---stretching, cutting, and painting---we enable the manipulation of logical qubits using strictly planar operations. We establish fault tolerance through numerical optimization of code distances and show that logical operations, including controlled-NOT gates, state transfers, and Pauli measurements, can be efficiently implemented within this framework to assemble an individually addressable logical qubit network. Universal quantum computation can then be realized by coupling just one BB-code logical qubit to a surface-code block. By combining the high encoding efficiency of qLDPC codes with geometric locality, our approach offers a practical and resource-efficient path to fault-tolerant quantum computation. 
\end{abstract}

\maketitle

\section*{Introduction}

One of the main obstacles to realizing fault-tolerant quantum computation is the large qubit overhead~\cite{Dennis2002,Bombin2006}. This challenge has motivated extensive efforts to search for quantum error correction codes with high encoding efficiency, particularly within the family of quantum low-density parity check (qLDPC) codes~\cite{Tillich2014,evra2022decodable,kaufman2021new,Hastings2021,Panteleev2021,breuckmann2021balanced,leverrier2022quantum,dinur2206good,panteleev2022asymptotically}. However, many such codes require long-range couplings between qubits, making them difficult to implement on two-dimensional platforms, for instance, superconducting quantum systems~\cite{Gottesman2013,Cohen2022,Xu2024}. A recent breakthrough is the discovery of bivariate bicycle (BB) codes~\cite{Bravyi2024,Voss2024,Wang2024,Wang2025,chen2025anyon,liang2025generalized}, along with a notable generalization to open boundaries~\cite{Steffan2025,Liang2025}, enabling low-overhead fault-tolerant quantum {\it memories} on planar hardware. 

In this work, we demonstrate that low-overhead fault-tolerant quantum {\it computation} can be realized on a planar qubit network. Two-dimensional topological codes, such as BB codes, are constructed by periodically tiling local stabilizers across a lattice. We define a {\it planar qubit network} as follows:
\begin{itemize}
\item An open-boundary two-dimensional array of qubits with strictly local connectivity---specifically, the couplings are tailored to accommodate the local stabilizers in the target code.
\end{itemize}
Enforcing open boundaries and locality constraints suppress long-range interactions and reduce wire crossings, thereby lowering experimental error rates~\cite{Liang2025,Steffan2025,mathews2025placing}. Many fault-tolerant protocols can, in principle, be embedded into such local-interaction qubit networks through the use of long-range communication~\cite{gottesman2000fault,svore2006noise,berthusen2024toward}. We separate our approach from these communication-based protocols by introducing the notion of {\it planar operations}, defined as: 
\begin{itemize}
\item Composite operations comprising exclusively two primitives: single-qubit operations and measurements of local stabilizers. Each stabilizer measurement is restricted to a single ancilla qubit interacting solely with the data qubits to which it is directly coupled.
\end{itemize}
Since such operations are free of communication overhead, they are more practical and tend to support higher fault-tolerance thresholds. For example, in surface-code schemes, logical controlled-NOT gates implemented via lattice surgery are planar operations~\cite{Horsman2012,Fowler2012,Litinski2019}. In addition to planarity, {\it translational symmetry} is another desirable feature, as it could simplify hardware design. Refer to the Supplementary Information for formal definitions of the planar qubit network, planar operations, and translational symmetry.
In this work, we employ high-encoding-rate planar BB codes to reduce logical qubit overhead, and we develop a systematic framework for designing the corresponding logical operations under strict planarity constraints.

\begin{figure}[htbp]
\centering
\includegraphics[width=\linewidth]{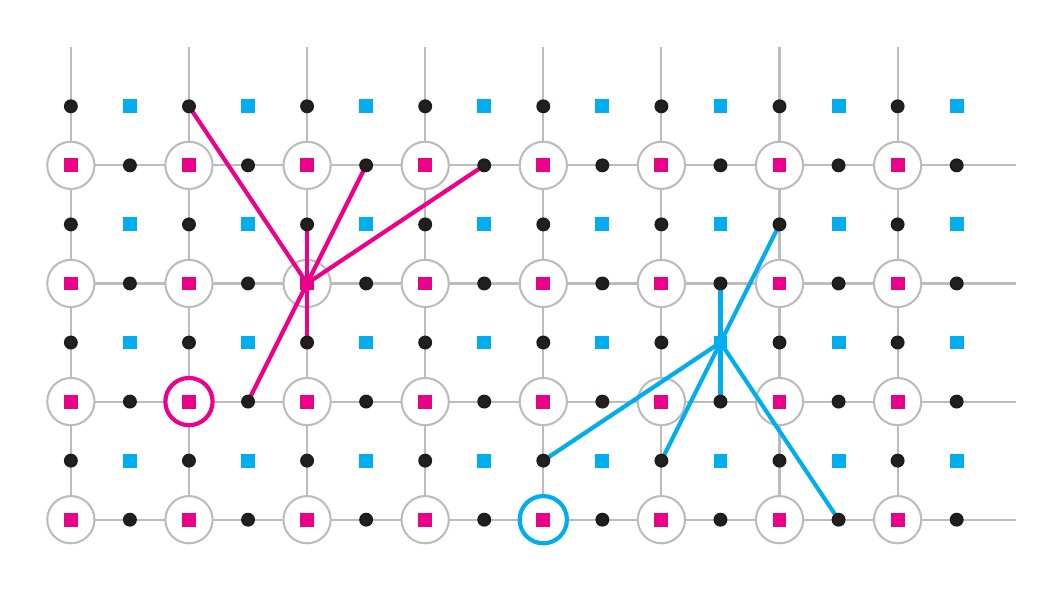}
\caption{
Planar qubit network tailored to stabilizers of the $[[54,6,4]]$ code~\cite{Liang2025}. Black filled circles represent data qubits, while red and blue squares denote ancilla qubits for measuring $X$ and $Z$ stabilizers, respectively, all arranged on an open-boundary two-dimensional square lattice. Two-qubit gates link these physical qubits into a network. Due to the translational symmetry, the coupling pattern is generated by repeating a fixed unit cell across the lattice. For clarity, only the couplings incident to representative red and blue ancilla qubits are shown (indicated by red and blue edges, respectively); all other ancilla qubits follow the same periodic coupling pattern with their nearby data qubits. Following the tile code convention~\cite{Steffan2025} (see Fig.~\ref{fig:code}), the ancilla qubits are spatially separated from the locations of stabilizers they measure; red and blue open circles indicate these stabilizer locations measured by the representative ancilla qubits.}
\label{fig:layout}
\end{figure}

We propose code craft as a systematic method for designing strictly planar operations on translationally invariant networks tailored to certain BB-type codes; see Fig.~\ref{fig:layout}. Our focus is on operations within the general framework of code surgery~\cite{Cohen2022,Cross2024,Cowtan2024,Ide2024,Williamson2024,Zhang2025,He2025,Cowtan2025}, which extends lattice surgery from surface codes to arbitrary qLDPC codes. A primary issue in code surgery is achieving individually addressable operations on logical qubits within a high-encoding-rate code block. Existing schemes typically address this by constructing a deformed code that couple an ancilla system to the support of the target logical operators. However, this support-based approach is incompatible with the constraints of planar operations, as it usually requires stabilizers beyond those supported by the qubit network. The key difficulty lies in the fact that, according to the constraints, the ancilla system can only be coupled to the boundary of the code block, and its stabilizer structure is limited to that of the original code. Code craft overcomes this by generating deformed codes through a sequence of deliberately arranged local modifications to the original code, ensuring that each intermediate code preserves planarity. 

It is essential to maintain fault tolerance during logical operations. To this end, we introduce a technique---logical operator painting---that optimizes planar operations to ensure a sufficiently large code distance throughout the operation. Moreover, we propose a method for optimizing the logical basis to facilitate the necessary logical operations while minimizing qubit overhead. 

The planar operations developed through code craft enable Pauli-operator measurements on logical qubits, which in turn allow for the implementation of logical controlled-NOT gates and state transfers. These operations establish a connected network of logical qubits. Building on this logical network, we can complete the universal gate set by adopting a hybrid strategy that incorporates a surface-code block~\cite{Xu2024,Stein2024,Yoder2025}. 

\section*{Results}

\begin{figure}[htbp]
\centering
\includegraphics[width=\linewidth]{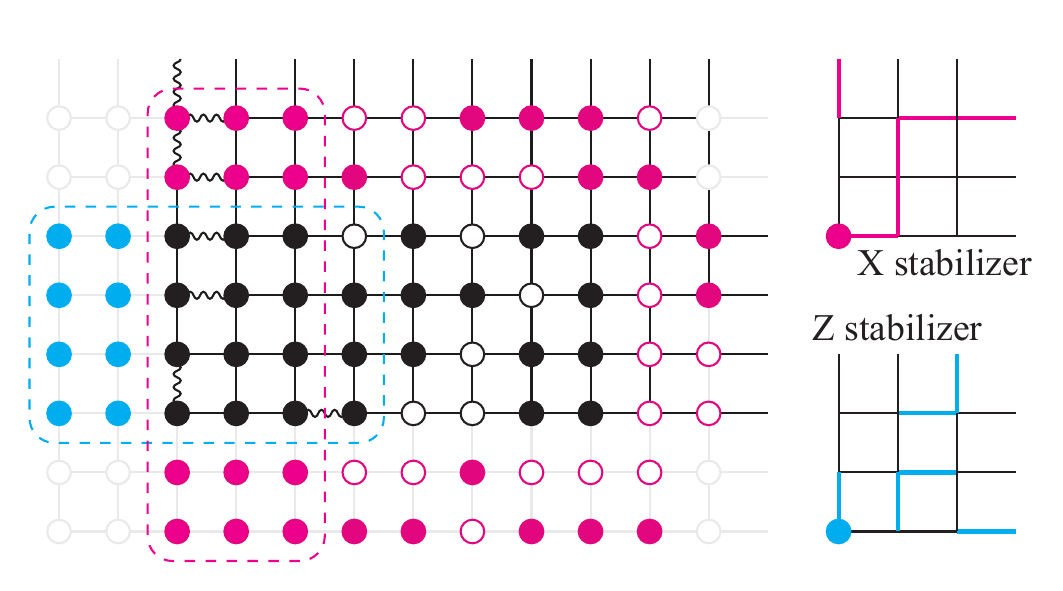}
\caption{
Code craft of the [[54,6,4]] code. Each edge represents a qubit, and each circle denotes a stabilizer generator: red and blue circles represent $X$-type and $Z$-type stabilizer generators, respectively, while black circles indicate overlapping $X$ and $Z$ generators at the same location. The structure of each stabilizer generator (up to qubit removal according to the cutting rule) is shown next to the qubit array, where red and blue edges indicate the qubits acted upon by $X$-type and $Z$-type generators, respectively. The full qubit array illustrates the intermediate code after applying stretching and $Z$ cutting; edges and circles in light gray indicate unused qubits and stabilizers. Dashed red and blue boxes outline the $X$ and $Z$ stabilizer generators of the original [[54,6,4]] code~\cite{Liang2025}, following the notation and conventions of Ref.~\cite{Steffan2025}. Wavy edges mark the support of an $X$ logical operator in the original code. To measure this operator, the $X$-type generators located at the open circles are removed during the $X$ cutting step. 
}
\label{fig:code}
\end{figure}

\subsection*{Planar BB codes and qubit networks} 
% \RED{We first describe the planar BB codes and qubit networks used in our construction.} 
Each family of BB-type codes is specified by two template stabilizers, as illustrated in Fig.~\ref{fig:code}~\cite{Bravyi2024}. A particular planar code is generated by periodically placing these template stabilizers across two overlapping regions on a square lattice. These two regions define the code~\cite{Steffan2025,Liang2025}.

To realize these codes, we arrange both data and ancilla qubits in a square lattice, as shown in Fig.~\ref{fig:layout}. Each stabilizer is assigned a dedicated ancilla qubit, and stabilizer measurements are implemented via direct couplings between ancilla qubits and their nearby data qubits. Consequently, the resulting qubit network naturally inherits the planarity and translational symmetry of the code structure.

\begin{figure}[htbp]
\centering
\includegraphics[width=\linewidth]{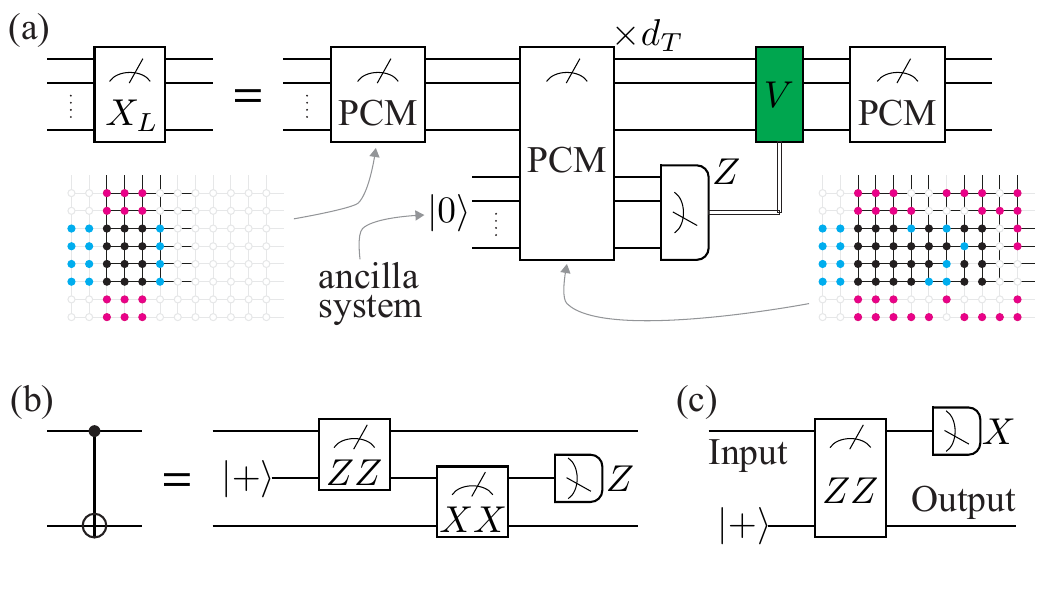}
\caption{
Circuits of logical operations. 
(a) Measurement of the logical operator $X_L$ (or a set of $X$ logical operators). Each horizontal line represents a physical qubit. Parity-check measurements (PCMs) are applied throughout the circuit to measure stabilizers. Two sets of stabilizers are involved: those defining the original code, measured once at the beginning and once at the end, and those defining the deformed code, measured repeatedly for $d_T$ rounds. Note that $V$ is a Pauli gate. 
(b) Implementation of a controlled-NOT gate between two logical qubits. 
(c) State transfer from one logical qubit to another. 
In (b) and (c), each horizontal line represents a logical qubit, and feedback gates depending on measurement outcomes have been omitted. 
}
\label{fig:circuit}
\end{figure}

\subsection*{Planar operations via code craft} 
% \RED{We next explain how planar operations are realized via code craft.}
A typical code surgery circuit is illustrated in Fig.~\ref{fig:circuit}(a). This circuit consists of single-qubit operations and stabilizer measurements performed on two codes: the original code and a deformed code. The circuit realizes a measurement of $X$ logical operators on the original code; a similar circuit can be used to measure $Z$ logical operators. By varying the deformed code, we can choose the operators to be measured. The logical measurements can then be used to implement logical gates; see Figs.~\ref{fig:circuit}(b)~and~(c) for examples. Next, we present the approach for designing deformed codes and optimizing the code surgery circuit. Given target logical operators, we can construct a dedicated deformed code from template stabilizers. As such, the code surgery circuit is a planar operation. 

\begin{figure}[htbp]
\centering
\includegraphics[width=\linewidth]{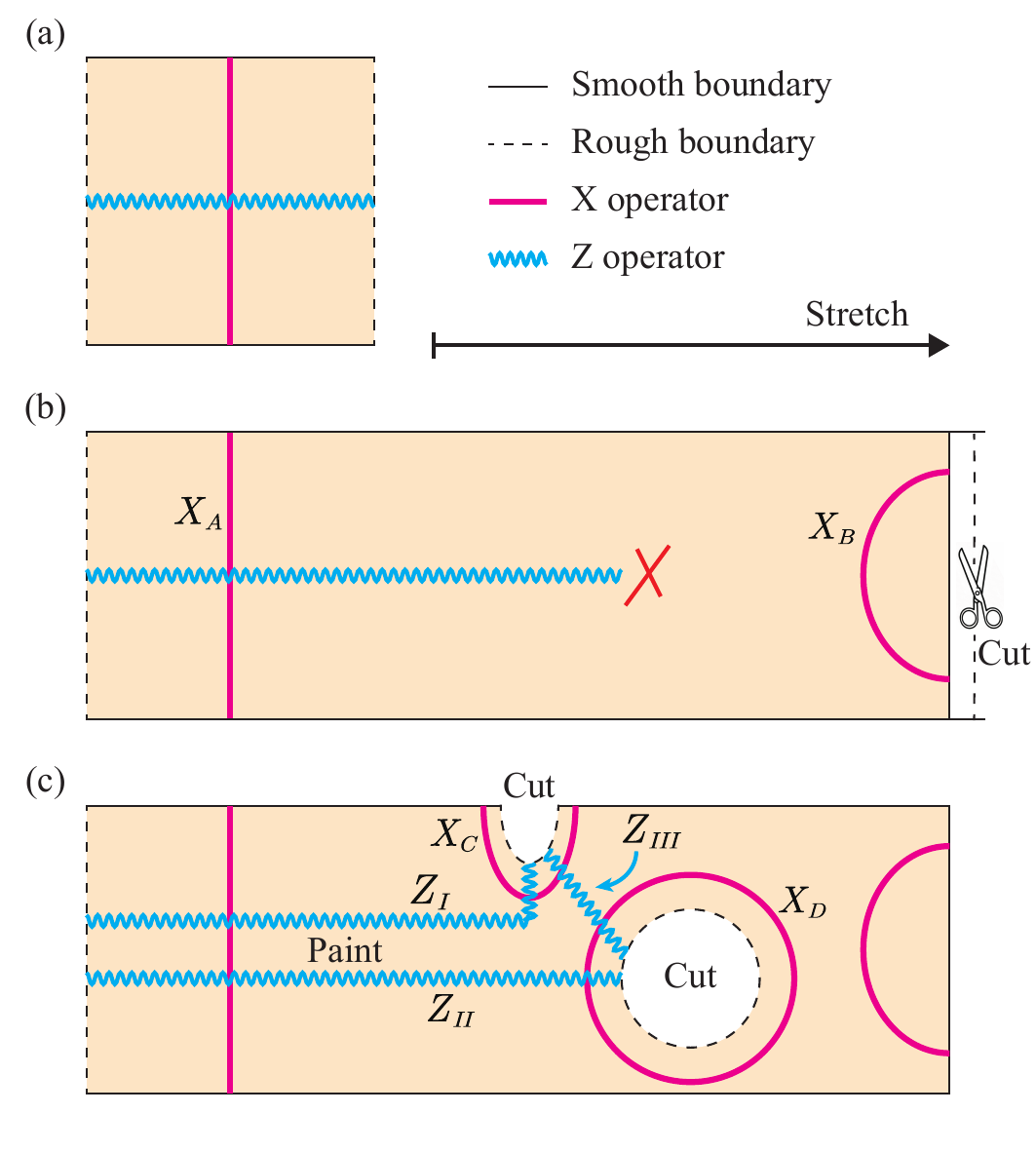}
\caption{
Schematic illustration of code craft. The actual $X$ cutting procedure removes $X$ stabilizers scattered across the lattice (see Fig.\ref{fig:code}), rather than forming regularly shaped geometric holes and boundaries as illustrated in (c). 
}
\label{fig:sketch}
\end{figure}

In code craft, we construct a deformed code dedicated to the target logical measurement through three operations on the original code: stretch, cut, and paint, as illustrated in Figs.~\ref{fig:code}~and~\ref{fig:sketch}. In what follows, we explain each of these operations taking the measurement of a single $X$ logical operator within one code block as an example. The method can then be generalized to $Z$ measurements, simultaneous measurements of multiple operators and inter-block measurements. 

%\textit{Lattice stretching.} 
\RED{The first code-craft operation is lattice stretching.} 
To explain this operation, we employ a picture from surface codes that reflects their topological nature~\cite{Dennis2002,Fowler2012}: a code block is a square with two types of boundaries, smooth and rough (also known as $m$-condensed and $e$-condensed boundaries, respectively), as shown in Fig.~\ref{fig:sketch}(a). Logical operators are visualized as lines connecting boundaries of the same type: a line terminating on smooth (rough) boundaries represents an $X$ ($Z$) logical operator. This smooth-rough-boundary picture extends naturally to planar BB codes, where the lines come in various species, each corresponding to a different logical qubit encoded within the block. 

To construct a code that measures an $X$ logical operator, we stretch the lattice by fixing one of the two rough boundaries and move the other to enlarge the code block. During this stretching process, the qubits introduced to fill the expanded region are collectively referred to as the ancilla system. In this step, the only subtlety lies in determining the appropriate length of the enlarged region, which we address through numerical analysis later. 

%\textit{Boundary $\boldsymbol{Z}$ stabilizer cutting.} Immediately following stretching, we remove certain $Z$-type stabilizers near the moved rough boundary. 
 \RED{The second code-craft operation is boundary $Z$ stabilizer cutting, in which we remove certain $Z$-type stabilizers near the moved rough boundary immediately after stretching.} This operation transforms the moved rough boundary into a smooth boundary. See Methods for details of designing the boundary. 

Stretching and $Z$ cutting create an enlarged code with a single smooth boundary covering three edges, as shown in Figs.~\ref{fig:code}~and~\ref{fig:sketch}(b). Let $X_A$ denote a logical operator of the original code. Since the code now has only one smooth boundary, $X_A$ is equivalent, by the topological nature of the code, to any $X$ operator that terminates on this boundary, such as $X_B$. This equivalence is mediated by an element of the stabilizer group, denoted $x(g_{AB})$, such that $X_A = x(g_{AB}) X_B$. By continuously deforming $X_B$, we can shrink its support until it becomes trivial, i.e.~$X_B = \openone$, which implies that $X_A$ itself is an element of the stabilizer group: $X_A = x(g_A)$. We refer to the product expression of $x(g_A)$ in terms of stabilizer generators as the measurement channel of $X_A$. 

We now define the measurement channel using the CSS formalism~\cite{Gottesman1997}. The original code before stretching is represented by two check matrices $H_X$ and $H_Z$, corresponding to $X$ and $Z$ stabilizers, respectively; and each of them has $n$ columns. Each $X$ logical operator of the original code is represented by an $n$-dimensional row vector $j_X$. The intermediate code created by stretching and $Z$ cutting has check matrices in the form~\cite{Ide2024,Williamson2024,Zhang20251}
\begin{eqnarray}
\bar{H}_X = \left(\begin{array}{cc}
H_X & 0 \\
S & H_G^\mathrm{T}
\end{array}\right)
\text{ and }
\bar{H}_Z = \left(\begin{array}{cc}
H_Z & T \\
0 & H_M
\end{array}\right). \notag
\end{eqnarray}
Here, the sub-matrices $H_G^\mathrm{T}$ and $H_M$ represent the $X$- and $Z$-type stabilizers of the ancilla system, respectively. The sub-matrices $S$ and $T$ describe the coupling between the ancilla system and the original code block. To ensure that all $X$ and $Z$ stabilizers commute, these matrices must satisfy the conditions $H_X S^\mathrm{T} = T H_G$ and $H_M H_G = 0$. In this formalism, a measurement channel of $j_X$ is a row vector $g(j_X)$ such that $(j_X, 0) = g(j_X) \bar{H}_X$. Note that each $X$ logical operator of the original code has a channel in the intermediate code. 

% \textit{Bulk $\boldsymbol{X}$ stabilizer cutting.} Next, we delete a set of selected $X$-type stabilizers from the intermediate code.
 
 \RED{The third code-craft operation is bulk $X$ stabilizer cutting, in which we delete a set of selected $X$-type stabilizers from the intermediate code.}
 The purpose is to close most measurement channels and only leave the one for the target $X$ logical operator. Specifically, if $j_{X,\mathrm{target}}$ is the $X$ logical operator to be measured, we remove all newly introduced $X$ stabilizers (i.e.,~the second row of $\bar{H}_X$) that lie outside the support of the channel $g(j_{X,\mathrm{target}})$; see Fig.~\ref{fig:code}. Suppose $g(j_{X,\mathrm{target}}) = (g_O,g_N)$, where $g_O$ and $g_N$ acts on the original and newly introduced $X$ stabilizers, respectively. On the check matrices, $X$ cutting modifies $\bar{H}_X$ by replacing the submatrices $S$ and $H_G^\mathrm{T}$ with $GS$ and $GH_G^\mathrm{T}$, respectively, where $G = \mathrm{diag}(g_N)$. 

We remark that after cutting operations, we always eliminate any qubits that are acted on by only one type of stabilizer; we refer to this as the cutting rule. As shown in~\cite{Steffan2025}, removing these qubits does not affect the commutativity of the stabilizers. This qubit removal can reduce both the qubit overhead and the potential errors associated with those qubits. 

After $X$ cutting and qubit removal, we obtain a deformed code capable of realizing the target logical measurement. Since the target $X$ logical operator is a product of stabilizers in the deformed code (due to the existence of the measurement channel), its eigenvalue can be read out through stabilizer measurements in the code surgery circuit. Meanwhile, other $X$ logical operators remain unmeasured, as their corresponding channels have been closed. 

The pseudocode for the $X$-cutting procedure and the generation of the corresponding deformed code is presented in Methods, accompanied by a formal proof (Theorem~1) establishing that the resulting code correctly implements the intended logical measurement. We emphasize, however, that this procedure relies on a verification step to identify potential instances where unintended logical operators are simultaneously measured alongside the target. While our numerical evaluations have not yet encountered such cases, we discuss in Supplementary Information how these occurrences could be resolved through a redefinition of the logical basis.

%\textit{Logical operator painting.} Finally, we address $Z$ logical operators during the measurement of an $X$ logical operator. 

\RED{The final code-craft operation is logical operator painting. It is used to address $Z$ logical operators during the measurement of an $X$ logical operator.} All $Z$ logical operators that commute with the measured $X$ operator must remain preserved throughout the code surgery circuit in Fig.~\ref{fig:circuit}(a). Consider such a commuting $Z$ logical operator, represented by an $n$-dimensional row vector $j_Z$. Its preservation proceeds as follows. First, after initializing qubits of the ancilla system in the $\ket{0}$ state, any operator of the form $\bar{j}_Z = (j_Z + h_Z, \beta)$ is equivalent to $j_Z$, where $h_Z \in \mathrm{rowsp}(H_Z)$ denotes an element in the stabilizer group of the original code and $\beta$ acts on the qubits initialized in $\ket{0}$. We refer to $(h_Z, \beta)$ as a valid storage of $j_Z$ if $\bar{j}_Z$ is a logical operator of the deformed code. Since $\bar{j}_Z$ commutes with subsequent deformed-code stabilizer measurements, it remains preserved until the ancilla system is measured out. Finally, we apply the feedback gate $V$ to correct the phase introduced by the storage and restore the original logical operator $j_Z$. We note that the deformed code constructed in the earlier stages guarantees valid storages for all $Z$ logical operators that commute with the measured $X$ operator. 

For each commuting $Z$ logical operator, there may exist multiple valid storages, which we will explain shortly. The choice of storage not only determines the form of the feedback gate $V$ but also affects the overall fault tolerance of the code surgery circuit. Therefore, we need to optimize the storage to ensure the fault tolerance. 

We now explain the storages of logical operators using the smooth-rough-boundary picture. Since the intermediate code produced by stretching and $Z$ cutting contains only one rough boundary, it does not support any $Z$ logical operators; see Fig.~\ref{fig:sketch}(b). The storages for $Z$ logical operators are created by $X$ cutting, which introduces additional rough boundaries in the code; see Fig.~\ref{fig:sketch}(c). Each $Z$ logical operator is then stored along a line terminating at the original fixed rough boundary and one of the newly created ones. These new rough boundaries also introduce gauge qubits, which result in multiple valid storages for a given $Z$ logical operator. For instance, $Z_I$ and $Z_{II}$ are both valid storages for a particular $Z$ logical operator, differing by a gauge operator $Z_{III}$. These different storages vary in their susceptibility to $X$ errors. For example, $Z_I$ and $Z_{II}$ anticommute with $X_C$ and $X_D$, respectively, and since $X_D$ has higher weight, we prefer $Z_{II}$ as a more fault-tolerant storage for the $Z$ logical operator of the original code. 

In logical operator painting, we optimize the storage of each $Z$ logical operator by adding appropriate gauge operators. An algorithm implementing this optimization is provided in Methods, which takes as input a target code distance $d_{\mathrm{th}}$ and an initial storage configuration. After optimization, it outputs a modified storage configuration such that the resulting deformed code, when treated as a subsystem code~\cite{Poulin2005,Kribs2005,Kribs2005a}, achieves a code distance of at least $d_{\mathrm{th}}$, up to the existence of a solution (Theorem~2). This guarantees that the code surgery circuit, provided that the deformed-code syndrome extraction is repeated for $d_T = \Theta(d_{\mathrm{th}})$ rounds, preserves the desired level of fault tolerance~\cite{Zhang20251,Williamson2024,Cross2024}.

This painting procedure transforms an initial subsystem code into a different subsystem code by redefining the logical operators, resulting in an enlarged code distance. We note that although the gauge operators of a subsystem code are typically not employed for detecting errors, certain gauge eigenvalues can be inferred from the final measurements on the ancilla system, as proposed in Ref.~\cite{Williamson2024}. Utilizing these gauge eigenvalues during the decoding process provides additional syndrome information, which potentially enhances the suppression of logical errors. \RED{A detailed comparison with the final-measurement approach of Ref.~\cite{Williamson2024} is provided in the Supplementary Information.}

While the operations of stretching, cutting, and painting are most intuitively visualized in the context of the surface code (Fig.~\ref{fig:sketch}), their application to general planar BB codes marks a transition from geometric to algebraic code deformation. In the surface code, bulk cutting removes a contiguous set of stabilizers to create a geometric void or hole [see Fig.~\ref{fig:sketch}(c)]. In general planar BB codes, however, the bulk-cutting step involves the strategic deletion of stabilizers---often scattered across the ancilla system---to algebraically isolate the target logical operator by eliminating the $k-1$ undesired operators from the measurement. The painting procedure serves as a numerical routine for restoring the code distance of the unmeasured logical operators after the cutting step; this is analogous to choosing a specific path for a logical operator in the surface-code picture. Consequently, our framework generalizes the geometric principles of traditional lattice surgery into an algebraic formalism, extending its utility to high-rate, multi-logical-qubit codes.

\begin{figure}[htbp]
\centering
\includegraphics[width=\linewidth]{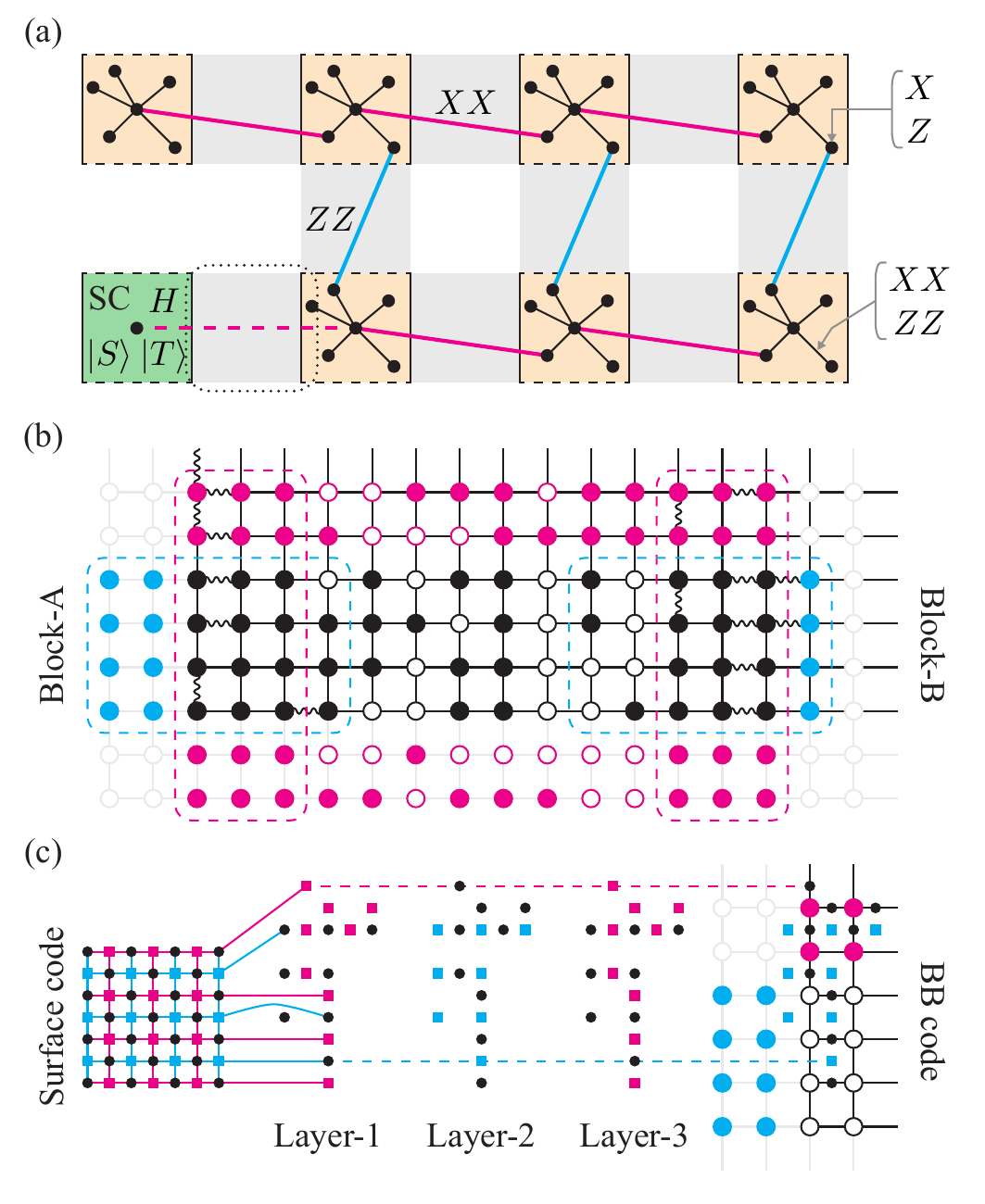}
\caption{
\RED{Universal quantum computation using a network of planar BB codes coupled to a surface-code block.}
(a) Array of code blocks. Orange squares denote planar-BB-code blocks, while the green square indicates a surface‑code block. Between these blocks are ancilla regions (gray squares), which are used to perform code surgery. Each black circle represents a logical qubit. Red and blue edges depict $XX$ and $ZZ$ joint measurements, respectively; black edges signify that both types of joint measurements can be performed between the connected qubits. 
(b) Code craft of the joint measurement on two blocks of the [[54,6,4]] code. 
(c) Deformed code used to implement the cross-code joint measurement indicated by the dashed edge in (a). As an example, we show the measurement between a distance-four surface code and one logical qubit from the [[54,6,4]] code block. The ancilla system [highlighted by the dotted box in (a)] consists of three layers of qubits, each replicating the layout and coupling pattern of a qubit subset from the BB-code region (black circles and blue squares). For visual clarity, the following are omitted: (i) intra-layer couplings within both the BB-code region and each ancilla layer, and (ii) transversal nearest-neighbor couplings between the ancilla layers and the corresponding BB-code qubits. Two representative transversal couplings between each pair of layers are shown as dashed edges, where each qubit is coupled to its counterpart in adjacent layers. 
}
\label{fig:scheme}
\end{figure}

\subsection*{Construction of logical networks} 
% \RED{We now describe the construction of logical networks.}
We have developed an approach for performing individually addressable logical measurements on a translationally invariant planar qubit network. 
These measurements serve as fundamental building blocks for constructing protocols for universal fault-tolerant quantum computation. In what follows, we present one such protocol. 

As a first step, we construct a network of logical qubits. As illustrated in Fig.~\ref{fig:scheme}(a), we consider a two-dimensional array of code blocks. For now, we focus on blocks of planar BB codes. A connected network of logical qubits can then be established as follows. 

\begin{table*}[t]
\centering
\small
% \begin{tabular}{cPPPPPPPPPPPPPPPPPPPPPP}
\setlength{\tabcolsep}{0pt}
\begin{tabular}{@{}c *{22}{c}@{}}

\hline\hline
\multicolumn{23}{c}{Single-block $X$ measurement} \\
~~Total ancilla size~~ & \multicolumn{2}{c}{$X_0$} & \multicolumn{2}{c}{$X_1$} & \multicolumn{2}{c}{$X_2$} & \multicolumn{2}{c}{$X_3$} & \multicolumn{2}{c}{$X_4$} & \multicolumn{2}{c}{$X_5$} & \multicolumn{2}{c}{$X_0X_1$} & \multicolumn{2}{c}{$X_0X_2$} & \multicolumn{2}{c}{$X_0X_3$} & \multicolumn{2}{c}{$X_0X_4$} & \multicolumn{2}{c}{$X_0X_5$} \\
\hline
32  & \bluecell{2} & \orangecell{5} & \bluecell{2} & \orangecell{4} & \bluecell{7} & \orangecell{7} & \bluecell{2} & \orangecell{6} & \bluecell{2} & \orangecell{7} & \bluecell{2} & \orangecell{5} & \bluecell{2} & \orangecell{8} & \bluecell{2} & \orangecell{7} & \bluecell{2} & \orangecell{6} & \bluecell{2} & \orangecell{6} & \bluecell{2} & \orangecell{5} \\
56  & \bluecell{2} & \orangecell{6} & \bluecell{2} & \orangecell{6} & \bluecell{9} & \orangecell{9} & \bluecell{2} & \orangecell{9} & \bluecell{2} & \orangecell{9} & \bluecell{2} & \orangecell{6} & \bluecell{2} & \orangecell{9} & \bluecell{5} & \orangecell{9} & \bluecell{2} & \orangecell{8} & \bluecell{2} & \orangecell{8} & \bluecell{2} & \orangecell{6} \\
80  & \bluecell{2} & \orangecell{7} & \bluecell{2} & \orangecell{9} &            &              &            &              &            &              & \bluecell{2} & \orangecell{6} &            &              &            &              & \bluecell{2} & \orangecell{9} & \bluecell{2} & \orangecell{9} & \bluecell{2} & \orangecell{8} \\
104 & \bluecell{2} & \orangecell{7} &            &              &            &              &            &              &            &              & \bluecell{2} & \orangecell{7} &            &              &            &              &            &              &            &              & \bluecell{2} & \orangecell{9} \\
128 & \bluecell{3} & \orangecell{8} &            &              &            &              &            &              &            &              & \bluecell{2} & \orangecell{9} &            &              &            &              &            &              &            &              &            &              \\
152 & \bluecell{2} & \orangecell{9} &            &              &            &              &            &              &            &              &            &              &            &              &            &              &            &              &            &              &            &              \\
\hline\hline
\multicolumn{23}{c}{Single-block $Z$ measurement} \\
~~Total ancilla size~~ & \multicolumn{2}{c}{$Z_0$} & \multicolumn{2}{c}{$Z_1$} & \multicolumn{2}{c}{$Z_2$} & \multicolumn{2}{c}{$Z_3$} & \multicolumn{2}{c}{$Z_4$} & \multicolumn{2}{c}{$Z_5$} & \multicolumn{2}{c}{$Z_0Z_1$} & \multicolumn{2}{c}{$Z_0Z_2$} & \multicolumn{2}{c}{$Z_0Z_3$} & \multicolumn{2}{c}{$Z_0Z_4$} & \multicolumn{2}{c}{$Z_0Z_5$} \\
\hline
27  & \bluecell{3} & \orangecell{5} & \bluecell{2} & \orangecell{5} & \bluecell{5} & \orangecell{5} & \bluecell{5} & \orangecell{5} & \bluecell{2} & \orangecell{5} & \bluecell{3} & \orangecell{5} & \bluecell{3} & \orangecell{4} & \bluecell{2} & \orangecell{6} & \bluecell{2} & \orangecell{5} & \bluecell{3} & \orangecell{7} & \bluecell{3} & \orangecell{5} \\
42  & \bluecell{2} & \orangecell{6} & \bluecell{3} & \orangecell{6} & \bluecell{6} & \orangecell{6} & \bluecell{6} & \orangecell{6} & \bluecell{4} & \orangecell{7} & \bluecell{4} & \orangecell{6} & \bluecell{4} & \orangecell{5} & \bluecell{3} & \orangecell{7} & \bluecell{3} & \orangecell{6} & \bluecell{4} & \orangecell{8} & \bluecell{2} & \orangecell{5} \\
57  & \bluecell{3} & \orangecell{7} & \bluecell{4} & \orangecell{7} & \bluecell{8} & \orangecell{8} & \bluecell{7} & \orangecell{7} & \bluecell{5} & \orangecell{8} & \bluecell{2} & \orangecell{8} & \bluecell{4} & \orangecell{6} & \bluecell{4} & \orangecell{8} & \bluecell{5} & \orangecell{7} & \bluecell{2} & \orangecell{9} & \bluecell{3} & \orangecell{6} \\
72  & \bluecell{3} & \orangecell{8} & \bluecell{2} & \orangecell{8} & \bluecell{9} & \orangecell{9} & \bluecell{9} & \orangecell{9} & \bluecell{3} & \orangecell{9} & \bluecell{3} & \orangecell{8} & \bluecell{3} & \orangecell{7} & \bluecell{2} & \orangecell{9} & \bluecell{6} & \orangecell{8} &            &              & \bluecell{3} & \orangecell{7} \\
87  & \bluecell{3} & \orangecell{9} & \bluecell{3} & \orangecell{9} &            &              &            &              &            &              & \bluecell{3} & \orangecell{9} & \bluecell{2} & \orangecell{9} &            &              & \bluecell{7} & \orangecell{8} &            &              & \bluecell{3} & \orangecell{8} \\
102 &            &              &            &              &            &              &            &              &            &              &            &              &            &              &            &              & \bluecell{7} & \orangecell{9} &            &              & \bluecell{3} & \orangecell{9} \\
% \end{tabular}

% \begin{tabular}{cPPPPPPPPPPPPPPPPPPPPPP}
% 替换原来的下两个表格部分
% \begin{tabular}{@{}c *{22}{c}@{}}
\hline\hline
\multicolumn{23}{c}{Two-block $X$ measurement} \\
~~Total ancilla size~~ 
& \multicolumn{2}{c}{$X_0 \otimes X_1$} 
& \multicolumn{2}{c}{\null} & \multicolumn{2}{c}{\null} & \multicolumn{2}{c}{\null}
& \multicolumn{2}{c}{\null} & \multicolumn{2}{c}{\null} & \multicolumn{2}{c}{\null}
& \multicolumn{2}{c}{\null} & \multicolumn{2}{c}{\null} & \multicolumn{2}{c}{\null} \\
\hline
180 
& \bluecell{2} & \orangecell{9} 
& \multicolumn{2}{c}{\null} & \multicolumn{2}{c}{\null} & \multicolumn{2}{c}{\null}
& \multicolumn{2}{c}{\null} & \multicolumn{2}{c}{\null} & \multicolumn{2}{c}{\null}
& \multicolumn{2}{c}{\null} & \multicolumn{2}{c}{\null} & \multicolumn{2}{c}{\null} \\
\hline\hline
\multicolumn{23}{c}{Two-block $Z$ measurement} \\
~~Total ancilla size~~ 
& \multicolumn{2}{c}{$Z_2 \otimes Z_3$} 
& \multicolumn{2}{c}{\null} & \multicolumn{2}{c}{\null} & \multicolumn{2}{c}{\null}
& \multicolumn{2}{c}{\null} & \multicolumn{2}{c}{\null} & \multicolumn{2}{c}{\null}
& \multicolumn{2}{c}{\null} & \multicolumn{2}{c}{\null} & \multicolumn{2}{c}{\null} \\
\hline
120 
& \bluecell{5} & \orangecell{9} 
& \multicolumn{2}{c}{\null} & \multicolumn{2}{c}{\null} & \multicolumn{2}{c}{\null}
& \multicolumn{2}{c}{\null} & \multicolumn{2}{c}{\null} & \multicolumn{2}{c}{\null}
& \multicolumn{2}{c}{\null} & \multicolumn{2}{c}{\null} & \multicolumn{2}{c}{\null} \\
\hline\hline
\end{tabular}

\caption{
Code distances of deformed codes for the [[180,6,9]] planar BB code, each corresponding to the measurement of a specific logical Pauli operator. Distances before and after logical operator painting are indicated in blue and orange, respectively. The total ancilla size refers to the number of data qubits within the ancilla system of the intermediate code, including both those actively participating in the logical-measurement circuit and those deactivated by the bulk-cutting operation. While distances for all two-qubit operators $X_a X_b$ and $Z_a Z_b$ are calculated, this table shows results only for the subset $X_0 X_b$ and $Z_0 Z_b$. Additional data, including results for [[54,6,4]] and [[162,8,7]] codes, are available in Supplementary Information. 
}
\label{table:180}
\end{table*}
For each code block, we can measure its $X$ and $Z$ logical operators using nearby ancilla regions. These logical measurements are carried out via code surgery, where the deformed code is generated using our code craft method. In generating the deformed code, the only parameter that needs to be determined is the stretching length. Additionally, we must optimize (through painting) and verify the distance of the deformed code. To this end, we conduct numerical simulations on three example planar BB codes: [[54,6,4]], [[180,6,9]], and [[162,8,7]]~\cite{Steffan2025,Liang2025}. The first two share the same template stabilizers but differ in size, while the third employs different template stabilizers. For each target logical operator, we compute the code distance of the corresponding deformed code. The results for the [[180,6,9]] code are shown in Table~\ref{table:180}, while those for the other two codes are provided in Supplementary Information, with consistent conclusions across all cases. From the numerical results, we observe that logical operator painting can significantly improve the code distance, and the optimized distance increases with the stretching length.

Regarding the necessary stretching length, i.e.~the minimal ancilla system size required to recover the original code distance, establishing a rigorous analytical bound remains an open challenge. Consequently, we characterize the properties of the deformed codes through numerical evaluation~\cite{cowtan2024ssip}. Our numerical simulations, conducted across a diverse set of codes with distances ranging from $d=5$ to $d=21$ (see Supplementary Information), demonstrate that the original code distance can always be restored. Furthermore, the necessary ancilla system size remains comparable to that of the original code block throughout this range, albeit with a trend of increasing relative overhead as the distance scales. This ensures the fault tolerance of the resulting logical measurements.

To construct a connected logical network, joint logical measurements between neighboring planar-BB-code blocks are also required. These measurements are implemented using the ancilla region located between the blocks. The corresponding deformed code is generated via code craft, with the key modification that the stretching and boundary $Z$ cutting steps are replaced by directly connecting the two code blocks, as illustrated in Fig.~\ref{fig:scheme}(b). Unlike single-block code craft, only a subset of $X$ logical operators admit measurement channels in this intermediate code. If the basis of the logical space (which defines logical qubits) is chosen arbitrarily, it may be the case that all allowed joint measurements act on multiple logical qubits within one code block, which is undesirable for constructing a logical network consisting of two-qubit operations. Therefore, it is essential to determine the logical basis in accordance with the measurement channels permitted by the intermediate code. 

In practice, we consider two types of intermediate codes, corresponding to horizontal and vertical couplings, respectively, to guide the choice of logical basis. Details of this procedure are provided in Supplementary Information. With this optimized basis, we find that inter-block logical measurements can be realized using ancilla systems only moderately larger than those used for single-block operations (see Table~\ref{table:180}), ultimately determining the required size of the ancilla regions between code blocks.

\subsection*{Universal quantum computation}
% \RED{We now explain how universal quantum computation is achieved.}
On the logical network, we already have a foundational set of operations: logical Pauli gates are straightforward to implement; single-qubit logical measurements enable initialization and readout; and two-qubit logical measurements allow for controlled-NOT gates and state transfers between any pair of planar-BB-code logical qubits. To complete the universal gate set, the remaining gates namely the Hadamard, $S$, and $T$ gates are provided through coupling to a surface-code block. 

At the logical level, this coupling requires only a single joint measurement between a planar-BB-code logical qubit and the surface-code logical qubit [the red dashed edge in Fig.~\ref{fig:scheme}(a)]. This measurement can be implemented using a deformed code constructed with various methods~\cite{Xu2024,Swaroop2024,Zhang20251}. Crucially, the use of only one such joint measurement allows us to tailor the physical qubit network to support just a single deformed code in the corresponding ancilla region, thereby significantly simplifying hardware requirements. The deformed code usually comprises the original BB and surface codes connected through a multi-layer ancilla system; see Fig.~\ref{fig:scheme}(c) for an example. Each ancilla layer can be viewed as a cropped portion of the BB-code block and requires no more than the same local couplings used in the BB code itself. These layers are arranged in a linear chain with transversal couplings, where the two end layers connect to the BB and surface codes, respectively. Consequently, the only additional hardware beyond the original BB and surface-code structures are the nearest-neighbor transversal connections between the ancilla layers and the adjacent code blocks. We note that these couplings break the translational symmetry in the region between the two codes, as indicated by the dotted box in Fig.~\ref{fig:scheme}(a). As an alternative to hardwiring the joint measurement into the physical layout, the surface code can be implemented on the BB-code qubit network by selectively deactivating certain couplings. Moreover, the nearest-neighbor transversal connections required for the joint measurement can be embedded within the BB-code qubit network using efficient routing protocols~\cite{berthusen2024toward}. This enables the joint measurement with the surface-code block to be executed entirely within the BB-code qubit network. 

With this joint measurement, we can transfer the state of a BB-code logical qubit to the surface code and implement the Hadamard gate there~\cite{Fowler2009,Horsman2012}. We can also prepare magic states~\cite{Li2014AMS,2024arXiv240303991I,Chamberland2020VeryLO,Bravyi2012,Wills2024,Gidney2024} on the surface code and transfer them to a BB-code logical qubit to realise $S$ and $T$ gates. Introducing multiple surface-code blocks can help reduce time overhead. In this way, the universal set of logical operations is completed. 

\section*{Discussion}

We have introduced code craft, a framework for designing logical operations on planar BB codes within planar physical qubit networks. Whereas many prior code surgery approaches focus on constructing deformed codes with provable code distances, our method uses numerical calculations to optimize (through logical operator painting) and verify fault tolerance with the goal of developing practical protocols. We remark that the core algorithms of code craft run in polynomial time with respect to code size, except for the subroutine used to evaluate code distance~\cite{landahl2011fault,chen2024error,pryadko2023qdistrnd}. Given that a moderate code distance is sufficient to suppress logical errors to acceptable levels for early quantum applications, this strategy provides a viable and efficient pathway for designing fault-tolerant protocols in near-term quantum devices. By leveraging the high encoding rates of the codes, our scheme enables fault-tolerant quantum computation on a planar architecture while significantly reducing overall qubit overhead. 

For future work, an important direction is the development of Hadamard gates and magic state injections implemented directly on planar BB codes. Achieving this will likely require extending beyond strictly planar operations to include certain quasi-planar operations, those that, while not strictly planar by definition, can still be efficiently realized on planar architectures. On surface codes, examples include Hadamard gates via lattice surgery and magic state cultivation~\cite{geher2024error,Fowler2012,Gidney2024,Chen2025}. While we conjecture that similar operations are feasible for planar BB codes, the hybrid structure employed in this work may offer practical advantages. In particular, surface codes have demonstrated strong performance for tasks such as magic state preparation. In this work, we have primarily focused on performing one operation at a time per block; however, code craft also supports parallel operations by manipulating measurement channels, which merits further exploration. Regarding the qubit overhead, it is worth noting that the translational invariance of the underlying physical lattice allows for the dynamic adjustment of inter-block spacing, which could minimize the total qubit cost and provide additional flexibility for inter-block logical operations. Furthermore, while stretching and boundary cutting are currently tailored to topological codes, the bulk cutting and painting operations are broadly applicable to general qLDPC codes. Specifically, given any deformed code that measures multiple logical operators, one can isolate a specific target operator by selectively deleting stabilizers and subsequently maximize the code distance by optimizing the logical operator storage. Consequently, our framework provides a versatile toolkit for constructing logical operations and enhancing fault tolerance.

\section*{Methods}

% \label{app:Zcut}

\subsection*{Boundary $Z$ stabilizer cutting} 
\begin{figure*}[htbp]
\centering
\includegraphics[width=\linewidth]{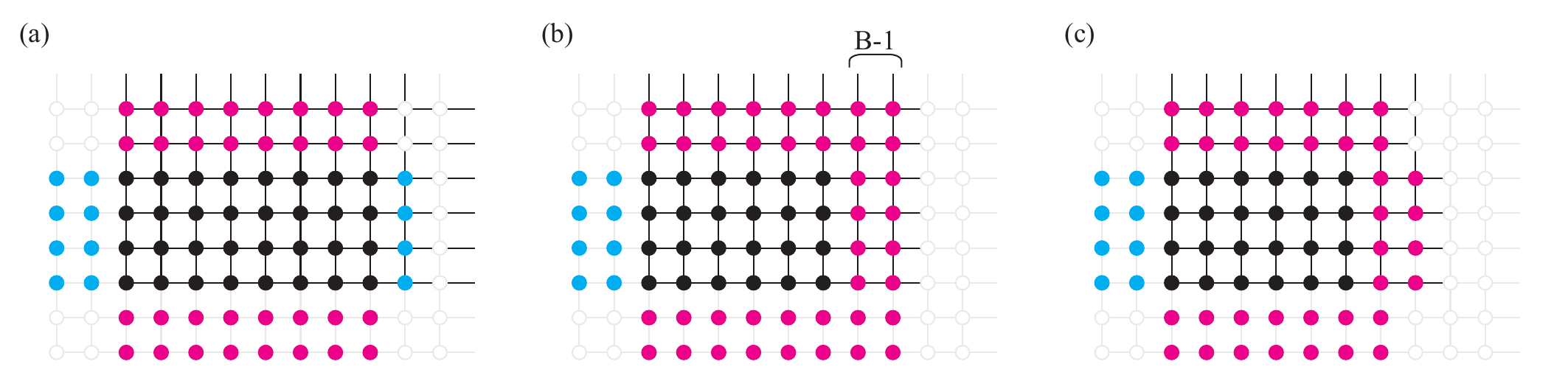}
\caption{
\RED{Boundary $Z$ stabilizer cutting procedure for the $[[54,6,4]]$ planar BB code.}
(a) An intermediate code generated by stretching the $[[54,6,4]]$ planar BB code. 
(b) The code after removing $Z$ stabilizers. For the $[[54,6,4]]$ planar BB code, the tile size is $B = 3$. 
(c) The eventual output code of $Z$ cutting. 
}
\label{fig:boundary}
\end{figure*}

% \RED{The boundary $Z$ stabilizer cutting procedure is as follows.}
For planar BB codes, smooth (rough) boundaries consists of boundary $X$ ($Z$) stabilizers indicated by red (blue) circles; black circles denote overlapped $X$ and $Z$ stabilizers, called bulk stabilizers; see Fig.~\ref{fig:boundary}(a). To construct a code that measures the $X$ logical operators, we convert one of the two rough boundaries into a smooth boundary. This is achieved by removing certain $Z$ stabilizers in the vicinity of that smooth boundary. This operation effectively reduces the number of logical qubits to zero. Now, we provide the detailed procedure: 
\begin{itemize}
\item As shown in Fig.~\ref{fig:boundary}(b), on the right-side rough boundary, remove boundary $Z$ stabilizers and $B-1$ columns of $Z$ stabilizers in the bulk region (converting $B-1$ columns of bulk stabilizers into boundary $X$ stabilizers), where $B$ is the size of template stabilizers (called tiles); 
\item Remove qubits according to the cutting rule, and remove all stabilizers whose supports are empty, resulting in the eventual code as shown in Fig.~\ref{fig:boundary}(c). 
\end{itemize}
Following the above procedure, we can generate a code with zero logical qubits that incorporates all the required measurement channels. We note that for certain codes, further fine-tuning near the boundary, referred to as topological-order completion~\cite{Liang2025}, could be necessary. 

% \label{app:Xcut}

\subsection*{Bulk $X$ stabilizer cutting}

% \RED{Next, we describe bulk $X$ stabilizer cutting.}
The pseudocode for the $X$ cutting procedure is provided in Algorithm~\ref{alg:Xcut}. 

\begin{theorem}
If the deformed code generated by Algorithm~\ref{alg:Xcut} passes the verification step, then it correctly implements the intended logical measurement. 
\end{theorem}

\begin{proof}
Without loss of generality, the check matrices $\tilde{H}_X$ and $\tilde{H}_Z$ can be written in the following form (up to permutations of rows and columns), 
\begin{eqnarray}
\tilde{H}_X &=& \left(\begin{array}{ccccc}
H_X & 0 & 0 & 0 & 0 \\
S' & H_G^{\prime \mathrm{T}} & 0 & B_X & 0 \\
0 & 0 & 0 & C_X & 0
\end{array}\right), \\
\tilde{H}_Z &=& \left(\begin{array}{ccccc}
H_Z & T' & A_Z & 0 & 0 \\
0 & H_M' & B_Z & 0 & 0  \\
0 & 0 & C_Z & 0 & 0
\end{array}\right).
\end{eqnarray}
Here, the third (fourth) block column corresponds to columns that are zero in $\tilde{H}_X$ ($\tilde{H}_Z$) but not in $\tilde{H}_Z$ ($\tilde{H}_X$). The fifth block column corresponds to columns that are zero in both $\tilde{H}_X$ and $\tilde{H}_Z$. The third block row corresponds to rows that are zero in the first two block columns. Since $H_X$ and $H_Z$ do not contain any zero columns or rows, the check matrices after deleting columns and rows read 
\begin{eqnarray}
\tilde{H}_X' &=& \left(\begin{array}{cc}
H_X & 0 \\
S' & H_G^{\prime \mathrm{T}}
\end{array}\right), \\
\tilde{H}_Z' &=& \left(\begin{array}{cc}
H_Z & T' \\
0 & H_M'
\end{array}\right).
\end{eqnarray}

From the structure of $\tilde{H}_X$ and $\tilde{H}_Z$, we identify 
\begin{eqnarray}
GS &=& \left(\begin{array}{c}
S' \\
0
\end{array}\right), \\
T &=& \left(\begin{array}{cccc}
T' & A_Z & 0 & 0
\end{array}\right), \\
G H_G^\mathrm{T} &=& \left(\begin{array}{cccc}
H_G^{\prime \mathrm{T}} & 0 & B_X & 0 \\
0 & 0 & C_X & 0
\end{array}\right), \\
H_M &=& \left(\begin{array}{cccc}
H_M' & B_Z & 0 & 0  \\
0 & C_Z & 0 & 0
\end{array}\right).
\end{eqnarray}
Given that $\bar{H}_X \bar{H}_Z^\mathrm{T} = 0$, it follows that $H_Z S^\mathrm{T} = T H_G$ and $H_M H_G = 0$. Multiplying both sides by $G^\mathrm{T}$ yields $H_ZS^\mathrm{T}G^\mathrm{T} = TH_GG^\mathrm{T}$ and $H_MH_GG^\mathrm{T} = 0$. By comparing both sides of these equations, we conclude that $H_ZS^{\prime \mathrm{T}} = T'H_G'$ and $H_M'H_G' = 0$. Hence, $\tilde{H}_X' \tilde{H}_Z^{\prime \mathrm{T}} = 0$, confirming that the check matrices $\tilde{H}_X'$ and $\tilde{H}_Z'$ define a valid CSS code. 

Recall the definition of $g(j_{X,\mathrm{target}})$, we have 
\begin{eqnarray}
\left(\begin{array}{cc}
j_{X,\mathrm{target}} & 0
\end{array}\right) = \left(\begin{array}{cc}
g_O H_X + g_N S & g_N H_G^\mathrm{T}
\end{array}\right).
\end{eqnarray}
Using $g_N G = g_N$, we have 
\begin{eqnarray}
\left(\begin{array}{cc}
j_{X,\mathrm{target}} & 0
\end{array}\right) = \left(\begin{array}{cc}
g_O H_X + g_N G S & g_N G H_G^\mathrm{T}
\end{array}\right).
\end{eqnarray}
Let us write $g_N = (g_N', g_N'')$, yielding 
\begin{eqnarray}
&&\left(\begin{array}{ccccc}
j_{X,\mathrm{target}} & 0 & 0 & 0 & 0
\end{array}\right) \notag \\
&=& \left(\begin{array}{ccccc}
g_O H_X + g_N' S' & g_N' H_G^{\prime \mathrm{T}} & 0 & g_N' B_X + g_N'' C_X & 0
\end{array}\right).~~~~~
\end{eqnarray}
Therefore, 
\begin{eqnarray}
\left(\begin{array}{ccccc}
j_{X,\mathrm{target}} & 0
\end{array}\right) &=& \left(\begin{array}{ccccc}
g_O H_X + g_N' S' & g_N' H_G^{\prime \mathrm{T}}
\end{array}\right) \notag \\
&=& \left(\begin{array}{ccccc}
g_O & g_N'
\end{array}\right) \tilde{H}_X',
\end{eqnarray}
which shows that $(j_{X,\mathrm{target}},0)$ lies in the row space of $\tilde{H}_X'$. Hence, the corresponding $X$ logical operator is an element of the stabilizer group of the deformed code and is therefore measured during the stabilizer measurements. 

If the verification step is passed, the vectors $(j_{X,a}', 0)$ are not in the row space of $\tilde{H}_X'$, implying that the corresponding $X$ logical operators are not part of the stabilizer group and are thus not measured. Furthermore, since $\tilde{H}_Z' (j_{X,a}', 0)^\mathrm{T} = 0$ holds for all $a$, these $X$ logical operators commute with all $Z$ stabilizers of the deformed code. Therefore, the unmeasured $X$ logical operators of the original code remain logical operators in the deformed code. This completes the proof. 
\end{proof}

\begin{figure*}[htbp]
\centering
\begin{minipage}{\textwidth}
\begin{algorithm}[H]
\caption{\textsc{$X$ stabilizer cutting} — Generation of the deformed code}
\label{alg:Xcut}
\begin{algorithmic}[1]
\State \textbf{Input:} Intermediate check matrices $\bar{H}_X$ and $\bar{H}_Z$ from stretching and $X$ cutting; independent $X$ logical operators $\{j_{X,1}, j_{X,2}, \ldots, j_{X,k}\}$ of the original code; target $X$ logical operator $j_{X,\mathrm{target}} \in \mathrm{Span}(j_{X,1}, \ldots, j_{X,k})$ to be measured. 
\State \textbf{Output:} Verified check matrices $\tilde{H}_X'$ and $\tilde{H}_Z'$ of the deformed code. 

\State Find $g(j_{X,\mathrm{target}})$ such that $(j_{X,\mathrm{target}}, 0) = g(j_{X,\mathrm{target}}) \bar{H}_X$.
\State Construct check matrices
\begin{eqnarray}
\tilde{H}_X = \left(\begin{array}{cc}
H_X & 0 \\
GS & GH_G^\mathrm{T}
\end{array}\right)
\text{ and }
\tilde{H}_Z = \left(\begin{array}{cc}
H_Z & T \\
0 & H_M
\end{array}\right), \notag
\end{eqnarray}
where $G = \mathrm{diag}(g_N)$ for $g(j_{X,\mathrm{target}}) = (g_O, g_N)$. 

\State Remove any zero columns in $\tilde{H}_X$ and the corresponding columns in $\tilde{H}_Z$; similarly, remove zero columns in $\tilde{H}_Z$ and the corresponding columns in $\tilde{H}_X$. 
\State Remove any zero rows in $\tilde{H}_X$ and $\tilde{H}_Z$. After removing the specified rows and columns, we obtain the eventual check matrices $\tilde{H}_X'$ and $\tilde{H}_Z'$. 

\State Find a basis of $\mathrm{Span}(j_{X,1}, \ldots, j_{X,k})$ containing $j_{X,\mathrm{target}}$, i.e., 
\[
\{j_{X,1}', j_{X,2}', \ldots, j_{X,k-1}', j_{X,\mathrm{target}}\}. 
\]
\State Verify that the rows of $\tilde{H}_X'$ together with $\{(j_{X,1}',0), \ldots, (j_{X,k-1}',0)\}$ are linearly independent. 
\Comment{If the vectors are linearly dependent, the deformed code will measure logical operators other than the target one.}
\end{algorithmic}
\end{algorithm}
\end{minipage}
\end{figure*}

% \label{app:paint}
\begin{figure*}[t]
\centering
\begin{minipage}{\textwidth}
\begin{algorithm}[H]
\caption{\textsc{Logical Operator Painting} – Storage Optimization}
\label{alg:paint}
\begin{algorithmic}[1]
\State \textbf{Input:} Independent $Z$ logical operators $\{j_{Z,1}, j_{Z,2}, \ldots, j_{Z,k}\}$ of the original code; unmeasured $X$ logical operators $\{j_{X,1}', j_{X,2}', \ldots, j_{X,k-1}'\}$ of the original code; check matrices $\tilde{H}_X'$ and $\tilde{H}_Z'$ of the deformed code adapted to the target operator; the code distance threshold $d_{\mathrm{th}}$. 
\State \textbf{Output:} A basis of commuting $Z$ logical operators $\{j_{Z,1}', j_{Z,2}', \ldots, j_{Z,k-1}'\}$ and their optimized storages $\{(h_{Z,1}, \beta_1), (h_{Z,2}, \beta_2), \ldots, (h_{Z,k-1}, \beta_{k-1})\}$. 

\State Construct a basis 
\[
\{u_1, \ldots, u_{k-1}\} \cup \{v_1, \ldots, v_m\} \cup \{\tilde{h}_{Z,1}, \ldots, \tilde{h}_{Z,r}\}
\]
of $\ker \tilde{H}_X'$ satisfying: 
\begin{itemize}
    \item[(i)] $(j_{X,a}', 0) u_b^\mathrm{T} = \delta_{a,b}$ for all $a,b = 1, 2, \ldots, k-1$; 
    \item[(ii)] $(j_{X,a}', 0) v_b^\mathrm{T} = 0$ for all $a=1, \ldots, k-1$ and $b=1, \ldots, m$; 
    \item[(iii)] $\{\tilde{h}_{Z,1}, \ldots, \tilde{h}_{Z,r}\}$ forms a basis of $\mathrm{rowsp}(\tilde{H}_Z')$. 
\end{itemize}

\For{$a = 1$ to $k-1$}
    \State $W \gets \{v_1, \ldots, v_m\}$ \Comment{Working set of vectors for updates}
    \State Find a vector $e$ of minimal Hamming weight satisfying: 
    \[
        \tilde{H}_Z' e^\mathrm{T} = 0, \quad u_a e^\mathrm{T} = 1
    \]
    \While{$|e| < d_{\mathrm{th}}$}
        \State Find $w \in W$ such that $w e^\mathrm{T} = 1$
        \If{no such $w$ exists}
            \State \textbf{Output:} Failure; \textbf{terminate}
        \EndIf
        \State Remove $w$ from $W$
        \ForAll{$w' \in W$ with $w' e^\mathrm{T} = 1$}
            \State $w' \gets w' + w$
        \EndFor
        \State Update $u_a \gets u_a + w$
        \State Recompute minimal-weight vector $e$ satisfying: 
        \[
            \tilde{H}_Z' e^\mathrm{T} = 0, \quad u_a e^\mathrm{T} = 1
        \]
    \EndWhile
\EndFor

\State Express the vectors $u_a$ in the form 
\[
u_a = (j_{Z,a}' + h_{Z,a}, \beta_a)
\]
with $j_{Z,a}' \in \mathrm{Span}(j_{Z,1}, j_{Z,2}, \ldots, j_{Z,k})$, yielding the basis of commuting $Z$ logical operators and their optimized storages. 
\end{algorithmic}
\end{algorithm}
\end{minipage}
\end{figure*}
\subsection*{Logical operator painting}
% \RED{Finally, we describe logical operator painting.}
The pseudocode for the painting procedure is provided in Algorithm~\ref{alg:paint}. 

\begin{theorem}
For every $Z$ logical operator that commutes with the measured $X$ logical operator, there exists a valid storage. Moreover, if there exists a valid storage configuration satisfying the distance threshold $d_{\mathrm{th}}$, then Algorithm~\ref{alg:paint} is guaranteed to find such a configuration and terminate successfully. 
\end{theorem}

\begin{proof}
First, we prove the existence of a valid storage. Since the rows of $\tilde{H}_X'$ together with $\{(j_{X,1}',0), \ldots, (j_{X,k-1}',0)\}$ are linearly independent, we can always find a basis of $\ker \tilde{H}_X'$ that satisfies the conditions specified in Line~3 of Algorithm~\ref{alg:paint}. Each vector $u_a$ in this basis corresponds to a $Z$ logical operator of the deformed code, and we write $u_a = (\alpha_a, \beta_a)$. From the condition $\tilde{H}_X' u_a^\mathrm{T} = 0$, it follows that $H_X \alpha_a^\mathrm{T} = 0$. Therefore, we can express $\alpha_a^\mathrm{T}$ as $j_{Z,a}' + h_{Z,a}$, where $j_{Z,a}'$ and $h_{Z,a}$ are a $Z$ logical operator and a $Z$ stabilizer operator of the original code, respectively. Consequently, $(h_{Z,a}, \beta_a)$ constitutes a storage for the logical operator $j_{Z,a}'$. 

Moreover, the conditions $j_{X,a}' j_{Z,b}^{\prime \mathrm{T}} = \delta_{a,b}$ and $j_{X,\mathrm{target}} j_{Z,b}^{\prime \mathrm{T}} = 0$ (since $(j_{X,\mathrm{target}}, 0)$ lies in the row space of $\tilde{H}_X'$) hold for all $a,b = 1, 2, \ldots, k-1$. Hence, the set $\{j_{Z,a}'\}$ forms a basis for the $Z$ logical operators of the original code that commute with the measured $X$ logical operator $j_{X,\mathrm{target}}$. Any such commuting $Z$ logical operator can be expressed as a linear combination of $j_{Z,a}'$, and the corresponding linear combination of their storages yields a valid storage for the given $Z$ logical operator. 

The above conclusions remain valid after the painting procedure. During painting, each $u_a$ vector is updated by adding vectors from $\mathrm{Span}(v_1, \ldots, v_m)$. Since each $v_b$ represents a $Z$ logical operator of the deformed code, the updated $u_a$ vectors continue to represent $Z$ logical operators. Furthermore, each $v_b$ satisfies $\tilde{H}_X' v_b^\mathrm{T} = 0$, which in particular implies $(j_{X,\mathrm{target}}, 0) v_b^\mathrm{T} = 0$. Thus, condition ii) in Line~3 of Algorithm~\ref{alg:paint} ensures that each $v_b$ can be written in the form $v_b = (h_{Z,b}', \beta_b')$, where $h_{Z,b}'$ is a $Z$ stabilizer operator of the original code. Therefore, at each step of painting, the updated $u_a$ continues to correspond to a valid storage for the same $Z$ logical operator $j_{Z,a}'$. 

Next, we prove the success of the algorithm, assuming the existence of a good storage configuration. Let $(h_a^\mathrm{g}, \beta_a^\mathrm{g})$ denote a good storage of the logical operator $j_{Z,a}'$, meaning that i) $u_a^\mathrm{g} = (j_{Z,a}' + h_a^\mathrm{g}, \beta_a^\mathrm{g})$ is a $Z$ logical operator of the deformed code, and ii) 
\begin{eqnarray}
\min\{ \vert e\vert ~:~ \tilde{H}_Z' e^\mathrm{T} = 0,~u_a^\mathrm{g} e^\mathrm{T} = 1\} \geq d_\mathrm{th}.
\end{eqnarray}
Let $u_a^\mathrm{i} = (j_{Z,a}' + h_a^\mathrm{i}, \beta_a^\mathrm{i})$ denote the initial storage operator generated in Line~3 of Algorithm\ref{alg:paint}. The good storage $u_a^\mathrm{g}$ can be obtained from $u_a^\mathrm{i}$ by adding the vector $\Delta = u_a^\mathrm{g} + u_a^\mathrm{i}$. Since both $u_a^\mathrm{i}$ and $u_a^\mathrm{g}$ are $Z$ logical operators of the deformed code, we can decompose $\Delta$ as $\Delta = \Delta_u + \Delta_v + \Delta_h$, where $\Delta_u \in \mathrm{Span}(u_1, \ldots, u_{k-1})$, $\Delta_v \in \mathrm{Span}(v_1, \ldots, v_m)$ and $\Delta_h \in \mathrm{Span}(\tilde{h}_{Z,1}, \ldots, \tilde{h}_{Z,r})$. 

Because $(j_{X,a}', 0)\Delta^\mathrm{T} = 0$ for all $a = 1, \ldots, k-1$, the component along $\mathrm{Span}(u_1, \ldots, u_{k-1})$ must vanish, i.e.~$\Delta_u = 0$. Moreover, adding a stabilizer operator $\Delta_h$ does not affect the code distance, so we may assume $\Delta_h = 0$ without loss of generality. Therefore, assuming the existence of a good storage $u_a^\mathrm{g}$, there exists a vector $\Delta_v \in \mathrm{Span}(v_1, \ldots, v_m)$ such that the $Z$ logical operator $u_a^\mathrm{g} = u_a^\mathrm{i} + \Delta_v$ achieves the required distance threshold. 

We prove the success of the painting procedure by contradiction. Suppose the painting of $u_a$ fails at the $t$-th iteration of the inner \texttt{while} loop in Algorithm~\ref{alg:paint} at Line 7. At the point of failure, the following vectors exist: 
\begin{enumerate}
\item A set of errors $\{e_s~:~s=1,\ldots,t\}$ with Hamming weight $\vert e_s\vert < d_\mathrm{th}$ for all $s$; 
\item A set of vectors $\{w_s~:~s=1,\ldots,t-1\}$ satisfying $w_se_s^\mathrm{T} = 1$ for all $s$; 
\item A set of vectors $\{w_s~:~s=t,\ldots,m\}$ satisfying $w_se_t^\mathrm{T} = 0$ for all $s$; 
\item An updated vector $u_a = u_a^\mathrm{i} + \sum_{s = 1}^{t-1} w_s$ satisfying $u_ae_s^\mathrm{T} = 0$ for all $s=1,\ldots,t-1$ and $u_ae_t^\mathrm{T} = 1$. 
\end{enumerate}
Here, $e_1$ is the error generated by Line~6 of the algorithm, and $e_{s>1}$ are generated at Line~15. The vectors $w_{s < t}$ are selected at Line~8, and the vectors $w_{s \geq t}$ are the remaining vectors in $W$ when the algorithm fails. Since $W$ is updated at each step (Lines~12 and 13), we also have the property 
\begin{itemize}
\item For each $s = 1, \ldots, t-1$ and all $s' > s$, $w_{s'} e_s^\mathrm{T} = 0$. 
\end{itemize}
Since $\{w_s~:~s=1,\ldots,m\}$ forms a basis for $\mathrm{Span}(v_1, \ldots, v_m)$, we can decompose $\Delta_v$ as $\Delta_v = \sum_{s = 1}^m \chi_s w_s$. Then $u_a^\mathrm{g} = u_a + \sum_{s = 1}^{t-1} (\chi_s+1) w_s + \sum_{s = t}^m \chi_s w_s$. If $\chi_s + 1 = 0$ for all $s = 1, \ldots, t-1$, then $u_a^\mathrm{g}e_t^\mathrm{T} = 1$, because $u_a e_t^\mathrm{T} = 1$ and $w_s e_t^\mathrm{T} = 0$ for $s \geq t$ by assumption. This contradicts the assumption that $u_a^\mathrm{g}$ has code distance at least $d_\mathrm{th}$ (since $e_t$ has weight below threshold and $u_a^\mathrm{g} e_t^\mathrm{T} = 1$). Therefore, at least one coefficient $\chi_q + 1$ (for $q < t$) must be nonzero. Let $q$ be the smallest such index. Then, $u_a^\mathrm{g}e_q^\mathrm{T} = 1$ (because $u_ae_q^\mathrm{T} = w_{s>q}e_q^\mathrm{T} = 0$), and again $\vert e_q \vert < d_\mathrm{th}$, contradicting the assumption that $u_a^\mathrm{g}$ satisfies the code distance condition. Hence, the assumption that painting fails leads to a contradiction. The algorithm must succeed if a good storage exists. 
\end{proof}

\section*{Data availability}
\RED{The numerical data for the code-distance analysis are available at \url{https://github.com/yyl19/Planar_BBcode}.}
\section*{Code availability}
\RED{The source codes for the numerical simulation are available at \url{https://github.com/yyl19/Planar_BBcode}. }
\section*{Acknowledgments}
This work is supported by the National Natural Science Foundation of China (Grant Nos. 12225507, 12088101) and NSAF (Grant No. U1930403). 

% \section*{Author contributions statement}
% All authors conceived the project and developed the protocol. Y.Y. performed the numerical simulations. All authors contributed to the analysis of the results and to writing the manuscript.
\section*{Author contributions}
\RED{Y.Y. initiated the project. Y.Y., G.Z., and Y.L. conceived the research and developed the protocol. Y.Y. conducted the numerical simulations. All authors analyzed the results and contributed to the preparation of the manuscript.}

\section*{Competing interests}
%The authors declare no competing interests. 
\RED{The authors declare no competing financial or non-financial interests.}

\bibliography{references.bib}
\clearpage
\section*{Figure legends}
\noindent\textbf{Figure 1}
Planar qubit network tailored to stabilizers of the $[[54,6,4]]$ code~\cite{Liang2025}. Black filled circles represent data qubits, while red and blue squares denote ancilla qubits for measuring $X$ and $Z$ stabilizers, respectively, all arranged on an open-boundary two-dimensional square lattice. Two-qubit gates link these physical qubits into a network. Due to the translational symmetry, the coupling pattern is generated by repeating a fixed unit cell across the lattice. For clarity, only the couplings incident to representative red and blue ancilla qubits are shown (indicated by red and blue edges, respectively); all other ancilla qubits follow the same periodic coupling pattern with their nearby data qubits. Following the tile code convention~\cite{Steffan2025} (see Fig.~\ref{fig:code}), the ancilla qubits are spatially separated from the locations of stabilizers they measure; red and blue open circles indicate these stabilizer locations measured by the representative ancilla qubits.

\vspace{0.5em}
\noindent\textbf{Figure 2}
Code craft of the [[54,6,4]] code. Each edge represents a qubit, and each circle denotes a stabilizer generator: red and blue circles represent $X$-type and $Z$-type stabilizer generators, respectively, while black circles indicate overlapping $X$ and $Z$ generators at the same location. The structure of each stabilizer generator (up to qubit removal according to the cutting rule) is shown next to the qubit array, where red and blue edges indicate the qubits acted upon by $X$-type and $Z$-type generators, respectively. The full qubit array illustrates the intermediate code after applying stretching and $Z$ cutting; edges and circles in light gray indicate unused qubits and stabilizers. Dashed red and blue boxes outline the $X$ and $Z$ stabilizer generators of the original [[54,6,4]] code~\cite{Liang2025}, following the notation and conventions of Ref.~\cite{Steffan2025}. Wavy edges mark the support of an $X$ logical operator in the original code. To measure this operator, the $X$-type generators located at the open circles are removed during the $X$ cutting step. 

\vspace{0.5em}
\noindent\textbf{Figure 3}
Circuits of logical operations. (a) Measurement of the logical operator $X_L$ (or a set of $X$ logical operators). Each horizontal line represents a physical qubit. Parity-check measurements (PCMs) are applied throughout the circuit to measure stabilizers. Two sets of stabilizers are involved: those defining the original code, measured once at the beginning and once at the end, and those defining the deformed code, measured repeatedly for $d_T$ rounds. Note that $V$ is a Pauli gate. 
(b) Implementation of a controlled-NOT gate between two logical qubits. 
(c) State transfer from one logical qubit to another. 
In (b) and (c), each horizontal line represents a logical qubit, and feedback gates depending on measurement outcomes have been omitted. 

\vspace{0.5em}
\noindent\textbf{Figure 4}
Schematic illustration of code craft. The actual $X$ cutting procedure removes $X$ stabilizers scattered across the lattice (see Fig.\ref{fig:code}), rather than forming regularly shaped geometric holes and boundaries as illustrated in (c). 

\vspace{0.5em}
\noindent\textbf{Figure 5}
\RED{Universal quantum computation using a network of planar BB codes coupled to a surface-code block.}
(a) Array of code blocks. Orange squares denote planar-BB-code blocks, while the green square indicates a surface‑code block. Between these blocks are ancilla regions (gray squares), which are used to perform code surgery. Each black circle represents a logical qubit. Red and blue edges depict $XX$ and $ZZ$ joint measurements, respectively; black edges signify that both types of joint measurements can be performed between the connected qubits. 
(b) Code craft of the joint measurement on two blocks of the [[54,6,4]] code. 
(c) Deformed code used to implement the cross-code joint measurement indicated by the dashed edge in (a). As an example, we show the measurement between a distance-four surface code and one logical qubit from the [[54,6,4]] code block. The ancilla system [highlighted by the dotted box in (a)] consists of three layers of qubits, each replicating the layout and coupling pattern of a qubit subset from the BB-code region (black circles and blue squares). For visual clarity, the following are omitted: (i) intra-layer couplings within both the BB-code region and each ancilla layer, and (ii) transversal nearest-neighbor couplings between the ancilla layers and the corresponding BB-code qubits. Two representative transversal couplings between each pair of layers are shown as dashed edges, where each qubit is coupled to its counterpart in adjacent layers. 

\vspace{0.5em}
\noindent\textbf{Figure 6}
\RED{Boundary $Z$ stabilizer cutting procedure for the $[[54,6,4]]$ planar BB code.}
(a) An intermediate code generated by stretching the $[[54,6,4]]$ planar BB code. 
(b) The code after removing $Z$ stabilizers. For the $[[54,6,4]]$ planar BB code, the tile size is $B = 3$. 
(c) The eventual output code of $Z$ cutting. 

\clearpage
\appendix
\setcounter{figure}{0}
\setcounter{table}{0}
\renewcommand{\figurename}{Supplementary Figure}
\renewcommand{\tablename}{Supplementary Table}
\makeatletter
\renewcommand{\fnum@figure}{Supplementary Figure~\thefigure}
\renewcommand{\fnum@table}{Supplementary Table~\thetable}
\@ifundefined{theHfigure}{}{\renewcommand{\theHfigure}{suppfigure.\arabic{figure}}}
\@ifundefined{theHtable}{}{\renewcommand{\theHtable}{supptable.\arabic{table}}}
\makeatother

\begin{widetext}
\section{Formal definitions}

\begin{definition}\textbf{Two-Dimensional Translationally Invariant Code.}
Let each site $\mathbf{r}$ of a square lattice $\mathbb{Z}^2$ contain a unit cell of $q$ qubits. The code is a stabilizer code whose group is generated by $c$ local templates shifted across the lattice. Specifically, if $\{g_{\mathbf{0},l}\}_{l=1}^c$ is the set of stabilizer generators centered at the origin, then the full set of generators is given by the orbits under the translation group:
\begin{equation}
g_{\mathbf{r},l} = T_{\mathbf{r}} (g_{\mathbf{0},l}), \quad \forall \mathbf{r} \in \mathbb{Z}^2, \, l \in \{1, \dots, c\}.
\end{equation}
where $T_{\mathbf{r}}$ is the translation operator acting on the qubits.
\end{definition}

This definition encompasses all BB-type codes considered in this work by setting $q = c = 2$. We note that an additional condition is required to justify topological order~\cite{haah2021classification}.

\begin{definition}\textbf{Planar Qubit Network.}
Given a two-dimensional translationally invariant code, the associated \textit{infinite planar qubit network} is defined by the Tanner graph of the code. Specifically, the network is embedded in a two-dimensional square lattice $\mathbb{Z}^2$, where each site $\mathbf{r}$ contains $q+c$ physical qubits. These qubits are partitioned into $q$ data qubits and $c$ ancilla qubits. The physical qubits form the vertices of an interaction graph. For each ancilla qubit, indexed as $(\mathbf{r}, A, l)$ for $l \in \{1, \dots, c\}$, the set of its adjacent vertices is identical to the support of the local stabilizer $g_{\mathbf{r},l}$. Two-qubit gates are restricted to pairs of adjacent qubits in this graph.

A \textit{finite planar qubit network} is defined as any subgraph of the infinite planar qubit network.
\end{definition}

\begin{definition}\textbf{Translational Symmetry of a Finite Qubit Network.}
A finite qubit network is said to possess translational symmetry if it is a subgraph of an infinite qubit network that is translationally symmetric.
\end{definition}

As an infinite planar qubit network is translationally symmetric by definition, the translational symmetry of its finite counterparts follows immediately.

\begin{definition}\textbf{Planar Operation.}
Consider a planar qubit network tailored to a two-dimensional translationally invariant code with local stabilizer generators $\{g_{\mathbf{r},l}\}$. An operation is defined as a planar operation if it comprises exclusively the following primitives:
\begin{enumerate}
\item Single-qubit operations;
\item Measurements of modified local stabilizers $\{g'_{\mathbf{r},l}\}$, where each $g'_{\mathbf{r},l}$ is the restriction of $g_{\mathbf{r},l}$ to a subset of its support. Formally, this implies $\mathrm{supp}(g'_{\mathbf{r},l}) \subseteq \mathrm{supp}(g_{\mathbf{r},l})$ and that $g'_{\mathbf{r},l}$ acts identically to $g_{\mathbf{r},l}$ on their common support.
\end{enumerate}
Furthermore, the measurement of each operator $g'_{\mathbf{r},l}$ must be implemented via the dedicated ancilla qubit $(\mathbf{r}, A, l)$ and its native couplings to adjacent data qubits.
\end{definition}

While not explicitly stated in the definition, it is an implicit requirement that the modified local stabilizers define a code with a sufficiently large distance to ensure fault tolerance. For example, the logical measurement circuit illustrated in Fig.~3 in the main text constitutes a planar operation, as it comprises a sequence of stabilizer measurements on codes sharing the same templates. In contrast, a standard surface-code Hadamard gate implemented on a two-dimensional qubit network is not strictly planar under our definition; such an implementation typically relies on SWAP gates to reposition code blocks~\cite{geher2024error}, which fall outside the restricted set of single-qubit operations and stabilizer measurements.

The above definitions are applicable to directional codes~\cite{geher2025directional}, notwithstanding their formulation in terms of ancilla qubits that shift positions during the stabilizer measurement cycle.

\section{Logical operator basis}
\label{app:basis}

In our protocol, the required ancilla system size depends on the specific logical operator being measured; additionally, for a fixed distance between two code blocks, only a specific subset of joint logical operators can be measured. The selection of the logical basis is therefore critical, as it determines both the eventual qubit overhead and the resulting topology of the logical qubit network. In this section, we describe how to choose a basis to avoid failures in the construction of single-block, single-logical-operator measurements, and we detail the approach used to determine the logical basis in our numerical simulations.

\subsection{Verification-step failures and logical basis redefinition}
\label{app:basis-redef}

In this section, we addresses the scenario flagged by the \emph{verification step} (Step~8) of Algorithm 1 in the main text, wherein the code deformation intended to measure a specific target logical operator (e.g.,~$X_1$) may inadvertently measure multiple logical operators simultaneously.

Our protocol involves two essential steps. First, by applying stretching and boundary cutting, we obtain a deformed code whose stabilizer group includes all $X$-type logical operators. Second, by applying bulk cutting, we remove $X$-type stabilizers on the ancilla system that are not involved in expressing the target operator $X_1$. Since $X_1$ can be written as a product of stabilizers in some subset $S_1$, we remove all ancilla-system $X$ stabilizers outside $S_1$. This ensures that $X_1$ remains measurable with the deformed code. However, this procedure does not guarantee that the deformed code measures only $X_1$: it may inadvertently measure additional logical operators as well. This is precisely the situation that triggers the ``verification step'' in Algorithm 1. Although our numerical studies do not exhibit such failure cases, we cannot rule them out in general.

Even in the event that a particular logical operator cannot be measured in isolation, this does not preclude constructing a complete logical operation set. Specifically, any $k$ independent $X$ logical operators may be designated as the single-logical-qubit $X$ operators, with the logical qubits defined accordingly. For example, consider the goal of constructing a set of logical operations that encompasses the measurement of each individual single-logical-qubit $X$ operator. Suppose the code deformation intended to isolate $X_1$ instead results in a measurement of both $X_1$ and $X_2$. Let $S_j$ denote the stabilizer support of the logical operator $X_j$, such that $X_j = \prod_{g \in S_j} g$. We partition $S_j$ into two subsets, $S_{o,j}$ and $S_{a,j}$, containing stabilizers from the original code and the ancilla system, respectively. The simultaneous measurement of $X_1$ and $X_2$ implies that $S_{a,2} \subseteq S_{a,1}$. To resolve the unintended simultaneous measurement, we can redefine the logical basis by taking $X_1' = X_1 X_2$ and $X_2' = X_2$, corresponding to ancilla-stabilizer supports $S_{a,1} - S_{a,2}$ and $S_{a,2}$, respectively. Crucially, both $S_{a,1} - S_{a,2}$ and $S_{a,2}$ must be non-empty; otherwise, $X_j'$ would reside in the original-code stabilizer group, contradicting the definition of a logical operator. By further removing ancilla-system stabilizers from one of these disjoint supports, we can isolate the measurement of the remaining redefined logical operator.

\subsection{A procedure for determining the logical basis}

The pseudocode for determining the logical basis is provided in Algorithm~\ref{alg:basis}, which is used in our numerical simulation.

\begin{algorithm}[H]
\caption{\textsc{Logical Basis Optimization}}
\label{alg:basis}
\begin{algorithmic}[1]
\State \textbf{Input:} Check matrices of the original code $H_X$ and $H_Z$; distances $s_{xx}$ and $s_{zz}$ characterizing the horizontal and vertical separation between two BB-code blocks. 
\State \textbf{Output:} $X$ logical operators $\{j_{X,1}, j_{X,2}, \ldots, j_{X,k}\}$ and $Z$ logical operators $\{j_{Z,1}, j_{Z,2}, \ldots, j_{Z,k}\}$ of the original code satisfying $j_{X,a}j_{Z,b}^\mathrm{T} = \delta_{a,b}$ for all $a,b = 1,2,\ldots,k$. 

\State Construct a deformed code by horizontally connecting two blocks of the original code separated by $s_{xx}$. The resulting check matrices are:
\begin{eqnarray}
\bar{H}_X^{(xx)} = \left(\begin{array}{ccc}
H_X & 0 & 0 \\
S_1 & H_1^\mathrm{T} & S_2 \\
0 & 0 & H_X
\end{array}\right)
\text{ and }
\bar{H}_Z^{(xx)} = \left(\begin{array}{ccc}
H_Z & T_1 & 0 \\
0 & H_2 & 0 \\
0 & T_2 & H_Z
\end{array}\right). \notag
\end{eqnarray}

\State Find a vector of the form $(j_{X,1}, 0, j_{X,2})$ such that: 
\begin{itemize}
    \item[(i)] $j_{X,1}, j_{X,2} \in \ker H_Z$; 
    \item[(ii)] $j_{X,1}, j_{X,2}$ and the rows of $H_X$ are linearly independent; 
    \item[(iii)] $(j_{X,1}, 0, j_{X,2}) \in \mathrm{rowsp}(\bar{H}_X^{(xx)})$. 
\end{itemize}
\Comment{If no such vector is found, try other values of $s_{xx}$.}

\State Construct a deformed code by vertically connecting two blocks of the original code separated by $s_{zz}$. The resulting check matrices are:
\begin{eqnarray}
\bar{H}_X^{(zz)} = \left(\begin{array}{ccc}
H_X & T_3 & 0 \\
0 & H_3 & 0 \\
0 & T_4 & H_X
\end{array}\right)
\text{ and }
\bar{H}_Z^{(xx)} = \left(\begin{array}{ccc}
H_Z & 0 & 0 \\
S_3 & H_4^\mathrm{T} & S_4 \\
0 & 0 & H_Z
\end{array}\right). \notag
\end{eqnarray}

\State Find a set of linearly independent vectors of the form
\[
A_l = (A_{L,l}, 0, A_{R,l})
\]
satisfying:
\begin{itemize}
    \item[(i)] $A_{L,l}, A_{R,l} \in \ker H_X$;
    \item[(ii)] $(A_{L,l}, 0, A_{R,l}) \in \mathrm{rowsp}(\bar{H}_Z^{(zz)})$.
\end{itemize}

\State Form a matrix $A$ with rows $A_l$, so that
\[
A = (A_L, 0, A_R).
\]

\State Solve the system
\[
\begin{cases}
q A_L j_{X,1}^\mathrm{T} = 0, \\
q A_L j_{X,2}^\mathrm{T} = 0, \\
q A_R j_{X,1}^\mathrm{T} = 0, \\
q A_R j_{X,2}^\mathrm{T} = 0,
\end{cases}
\]
and choose a solution $q$ such that the vectors $q A_L$, $q A_R$, and the rows of $H_Z$ are linearly independent. 
\Comment{If no suitable solution is found, try different $(j_{X,1}, 0, j_{X,2})$ or vary $s_{xx}$ and $s_{zz}$.}

\State Set $j_{Z,3} \gets q A_L$ and $j_{Z,4} \gets q A_R$. 

\State Complete the basis construction for the logical operators. 

\end{algorithmic}
\end{algorithm}

\section{Additional numerical results}

\subsection{Distance data for three example codes}
\label{app:data}

Data are provided in Tables~\ref{table:180II}, \ref{table:54}, and \ref{table:162}. 

\begin{table*}[t]
\centering
\small
\begin{tabular}{cPPPPPPPPPPPPPPPPPPPPPP}
\hline\hline
\multicolumn{23}{c}{Single-block $X$ measurement} \\
Total ancilla size & \multicolumn{2}{c}{$X_1X_2$} & \multicolumn{2}{c}{$X_1X_3$} & \multicolumn{2}{c}{$X_1X_4$} & \multicolumn{2}{c}{$X_1X_5$}
             & \multicolumn{2}{c}{$X_2X_3$} & \multicolumn{2}{c}{$X_2X_4$} & \multicolumn{2}{c}{$X_2X_5$}
             & \multicolumn{2}{c}{$X_3X_4$} & \multicolumn{2}{c}{$X_3X_5$} & \multicolumn{2}{c}{$X_4X_5$} \\
%\hline
32 & \bluecell{2} & \orangecell{6} & \bluecell{3} & \orangecell{6} & \bluecell{2} & \orangecell{5} & \bluecell{2} & \orangecell{5} & \bluecell{2} & \orangecell{7} & \bluecell{3} & \orangecell{7} & \bluecell{2} & \orangecell{5} & \bluecell{6} & \orangecell{6} & \bluecell{2} & \orangecell{5} & \bluecell{2} & \orangecell{5} \\
56 & \bluecell{2} & \orangecell{8} & \bluecell{6} & \orangecell{8} & \bluecell{2} & \orangecell{7} & \bluecell{2} & \orangecell{7} & \bluecell{2} & \orangecell{9} & \bluecell{3} & \orangecell{9} & \bluecell{2} & \orangecell{6} & \bluecell{7} & \orangecell{7} & \bluecell{2} & \orangecell{8} & \bluecell{2} & \orangecell{7} \\
80 & \bluecell{2} & \orangecell{9} & \bluecell{9} & \orangecell{9} & \bluecell{2} & \orangecell{8} & \bluecell{2} & \orangecell{8} &  &  &  &  & \bluecell{3} & \orangecell{9} & \bluecell{2} & \orangecell{8} & \bluecell{2} & \orangecell{9} & \bluecell{2} & \orangecell{7}  \\
104 &  &  &  &  & \bluecell{2} & \orangecell{9} & \bluecell{2} & \orangecell{9} &  &  &  &  &  &  & \bluecell{2} & \orangecell{8} &  &  & \bluecell{3} & \orangecell{8} \\
128 &  &  &  &  &  &  &  &  &  &  &  &  &  &  & \bluecell{2}  & \orangecell{9} & &  & \bluecell{2}  & \orangecell{9}  &  & \\

\hline\hline
\multicolumn{23}{c}{Single-block $Z$ measurement} \\
Total ancilla size & \multicolumn{2}{c}{$Z_1Z_2$} & \multicolumn{2}{c}{$Z_1Z_3$} & \multicolumn{2}{c}{$Z_1Z_4$} & \multicolumn{2}{c}{$Z_1Z_5$}
             & \multicolumn{2}{c}{$Z_2Z_3$} & \multicolumn{2}{c}{$Z_2Z_4$} & \multicolumn{2}{c}{$Z_2Z_5$}
             & \multicolumn{2}{c}{$Z_3Z_4$} & \multicolumn{2}{c}{$Z_3Z_5$} & \multicolumn{2}{c}{$Z_4Z_5$} \\
%\hline
27 & \bluecell{4} & \orangecell{5} & \bluecell{4} & \orangecell{5} & \bluecell{4} & \orangecell{5} & \bluecell{3} & \orangecell{6} & \bluecell{4} & \orangecell{5} & \bluecell{3} & \orangecell{6} & \bluecell{3} & \orangecell{4} & \bluecell{3} & \orangecell{4} & \bluecell{4} & \orangecell{4} & \bluecell{4} & \orangecell{5} \\
42 & \bluecell{5} & \orangecell{6} & \bluecell{3} & \orangecell{6} & \bluecell{5} & \orangecell{5} & \bluecell{4} & \orangecell{6} & \bluecell{6} & \orangecell{7} & \bluecell{5} & \orangecell{7} & \bluecell{4} & \orangecell{5} & \bluecell{5} & \orangecell{5} & \bluecell{2} & \orangecell{5} & \bluecell{4} & \orangecell{6} \\
57 & \bluecell{6} & \orangecell{6} & \bluecell{6} & \orangecell{6} & \bluecell{6} & \orangecell{7} & \bluecell{2} & \orangecell{7} & \bluecell{7} & \orangecell{8} & \bluecell{5} & \orangecell{8} & \bluecell{2} & \orangecell{6} & \bluecell{2} & \orangecell{6} & \bluecell{3} & \orangecell{6} & \bluecell{3} & \orangecell{7} \\
72 & \bluecell{7} & \orangecell{7} & \bluecell{2} & \orangecell{7} & \bluecell{7} & \orangecell{8} & \bluecell{4} & \orangecell{9} & \bluecell{9} & \orangecell{9} & \bluecell{5} & \orangecell{8} & \bluecell{4} & \orangecell{7} & \bluecell{3} & \orangecell{8} & \bluecell{4} & \orangecell{6} & \bluecell{2} & \orangecell{8} \\
87 & \bluecell{9} & \orangecell{9} & \bluecell{4} & \orangecell{9} & \bluecell{9} & \orangecell{9} &  &  &  & & \bluecell{5} & \orangecell{9} & \bluecell{4} & \orangecell{8} & \bluecell{4} & \orangecell{8} & \bluecell{4} & \orangecell{6} &  \bluecell{3} & \orangecell{9} \\
102 &  &  &  &  &  &  &  &  &  &  &  &  & \bluecell{4} & \orangecell{9} & \bluecell{5} & \orangecell{8} &  \bluecell{4} & \orangecell{8}  &  &  \\
117 &  &  &  &  &  &  &  &  &  &  &  &  &  &  & \bluecell{5} & \orangecell{8} & \bluecell{4} & \orangecell{8} & &  \\
132 &  &  &  &  &  &  &  &  &  &  &  &  &  &  &\bluecell{5}  & \orangecell{9} & \bluecell{4} & \orangecell{9} &  &  \\

\hline
\end{tabular}
\caption{Supplementary data on code distances for the [[180,6,9]] BB code. 
}
\label{table:180II}
\end{table*}

\begin{table*}[t]
\centering
\small
\begin{tabular}{cPPPPPPPPPPPPPPPPPPPPPP}
\hline\hline
\multicolumn{23}{c}{Single-block $X$ measurement} \\
%\hline
Total ancilla size & \multicolumn{2}{c}{$X_0$} & \multicolumn{2}{c}{$X_1$} & \multicolumn{2}{c}{$X_2$} & \multicolumn{2}{c}{$X_3$} & \multicolumn{2}{c}{$X_4$} & \multicolumn{2}{c}{$X_5$} & \multicolumn{2}{c}{$X_0X_1$} & \multicolumn{2}{c}{$X_0X_2$} & \multicolumn{2}{c}{$X_0X_3$} & \multicolumn{2}{c}{$X_0X_4$} & \multicolumn{2}{c}{$X_0X_5$} \\
%\hline
14 & \bluecell{2} & \orangecell{2} & \bluecell{3} & \orangecell{3} & \bluecell{2} & \orangecell{3} & \bluecell{2} & \orangecell{2} & \bluecell{2} & \orangecell{4} & \bluecell{2} & \orangecell{3} & \bluecell{2} & \orangecell{2} & \bluecell{3} & \orangecell{4} & \bluecell{2} & \orangecell{4} & \bluecell{2} & \orangecell{2} & \bluecell{2} & \orangecell{3} \\
26 & \bluecell{2} & \orangecell{4} & \bluecell{2} & \orangecell{3} & \bluecell{2} & \orangecell{4} & \bluecell{2} & \orangecell{3} &  &  & \bluecell{2} & \orangecell{4} & \bluecell{2} & \orangecell{3} &  &  &  &  & \bluecell{2} & \orangecell{4} & \bluecell{2} & \orangecell{3} \\
38 &  &  & \bluecell{2} & \orangecell{4} &  &  & \bluecell{2} & \orangecell{3} &  &  &  &  & \bluecell{2} & \orangecell{4} &  &  &  &  &  &  & \bluecell{2} & \orangecell{3} \\
50 &  &  &  &  &  &  & \bluecell{2} & \orangecell{4} &  &  &  &  &  &  &  &  &  &  &  &  & \bluecell{2} & \orangecell{4} \\

%\hline
  & \multicolumn{2}{c}{$X_1X_2$} & \multicolumn{2}{c}{$X_1X_3$} & \multicolumn{2}{c}{$X_1X_4$} & \multicolumn{2}{c}{$X_1X_5$}
             & \multicolumn{2}{c}{$X_2X_3$} & \multicolumn{2}{c}{$X_2X_4$} & \multicolumn{2}{c}{$X_2X_5$}
             & \multicolumn{2}{c}{$X_3X_4$} & \multicolumn{2}{c}{$X_3X_5$} & \multicolumn{2}{c}{$X_4X_5$} \\
%\hline
14 & \bluecell{3} & \orangecell{4} & \bluecell{2} & \orangecell{3} & \bluecell{2} & \orangecell{3} & \bluecell{2} & \orangecell{3} & \bluecell{2} & \orangecell{3} & \bluecell{2} & \orangecell{2} & \bluecell{2} & \orangecell{3} & \bluecell{2} & \orangecell{2} & \bluecell{2} & \orangecell{2} & \bluecell{2} & \orangecell{3} \\
26 &  &  & \bluecell{2} & \orangecell{4} & \bluecell{2} & \orangecell{4} & \bluecell{3} & \orangecell{3} & \bluecell{2} & \orangecell{4} & \bluecell{2} & \orangecell{4} & \bluecell{2} & \orangecell{4} & \bluecell{2} & \orangecell{3} & \bluecell{2} & \orangecell{4} & \bluecell{2} & \orangecell{3} \\
38 &  &  &  &  &  &  & \bluecell{3} & \orangecell{3} &  &  &  &  &  &  & \bluecell{2} & \orangecell{4} &  &  & \bluecell{2} & \orangecell{4} \\
50 &  &  &  &  &  &  & \bluecell{3} & \orangecell{4} &  &  &  &  &  &  &  &  &  &  &  &  \\

\hline\hline
\multicolumn{23}{c}{Single-block $Z$ measurement} \\
%\hline
Total ancilla size & \multicolumn{2}{c}{$Z_0$} & \multicolumn{2}{c}{$Z_1$} & \multicolumn{2}{c}{$Z_2$} & \multicolumn{2}{c}{$Z_3$} & \multicolumn{2}{c}{$Z_4$} & \multicolumn{2}{c}{$Z_5$} & \multicolumn{2}{c}{$Z_0Z_1$} & \multicolumn{2}{c}{$Z_0Z_2$} & \multicolumn{2}{c}{$Z_0Z_3$} & \multicolumn{2}{c}{$Z_0Z_4$} & \multicolumn{2}{c}{$Z_0Z_5$} \\
%\hline
15 & \bluecell{4} & \orangecell{4} & \bluecell{4} & \orangecell{4} & \bluecell{3} & \orangecell{3} & \bluecell{2} & \orangecell{2} & \bluecell{4} & \orangecell{4} & \bluecell{3} & \orangecell{3} & \bluecell{2} & \orangecell{2} & \bluecell{3} & \orangecell{3} & \bluecell{3} & \orangecell{3} & \bluecell{2} & \orangecell{2} & \bluecell{3} & \orangecell{4} \\
24 &  &  &  &  & \bluecell{4} & \orangecell{4} & \bluecell{4} & \orangecell{4} &  &  & \bluecell{3} & \orangecell{3} & \bluecell{2} & \orangecell{2} & \bluecell{2} & \orangecell{4} & \bluecell{2} & \orangecell{4} & \bluecell{2} & \orangecell{3} &  &  \\
33 &  &  &  &  &  &  &  &  &  &  & \bluecell{3} & \orangecell{4} & \bluecell{3} & \orangecell{3} &  &  &  &  & \bluecell{3} & \orangecell{4} &  &  \\
42 &  &  &  &  &  &  &  &  &  &  &  &  & \bluecell{4} & \orangecell{4} &  &  &  &  &  &  &  &  \\
%\hline
  & \multicolumn{2}{c}{$Z_1Z_2$} & \multicolumn{2}{c}{$Z_1Z_3$} & \multicolumn{2}{c}{$Z_1Z_4$} & \multicolumn{2}{c}{$Z_1Z_5$}
             & \multicolumn{2}{c}{$Z_2Z_3$} & \multicolumn{2}{c}{$Z_2Z_4$} & \multicolumn{2}{c}{$Z_2Z_5$}
             & \multicolumn{2}{c}{$Z_3Z_4$} & \multicolumn{2}{c}{$Z_3Z_5$} & \multicolumn{2}{c}{$Z_4Z_5$} \\
%\hline
5 & \bluecell{3} & \orangecell{3} & \bluecell{2} & \orangecell{2} & \bluecell{4} & \orangecell{4} & \bluecell{3} & \orangecell{3} & \bluecell{4} & \orangecell{4} & \bluecell{4} & \orangecell{4} & \bluecell{3} & \orangecell{4} & \bluecell{2} & \orangecell{2} & \bluecell{3} & \orangecell{4} & \bluecell{3} & \orangecell{3} &  &  \\
24 & \bluecell{4} & \orangecell{4} & \bluecell{4} & \orangecell{4} &  &  & \bluecell{2} & \orangecell{3} &  &  &  &  &  &  & \bluecell{2} & \orangecell{2} &  &  & \bluecell{2} & \orangecell{3} &  &  \\
33 &  &  &  &  &  &  & \bluecell{3} & \orangecell{4} &  &  &  &  &  &  & \bluecell{3} & \orangecell{3} &  &  & \bluecell{3} & \orangecell{4} &  &  \\
42 &  &  &  &  &  &  &  &  &  &  &  &  &  &  & \bluecell{2} & \orangecell{4} &  &  &  &  &  &  \\

\hline\hline
\end{tabular}

\begin{tabular}{cPPPPPPPPPPPPPPPPPPPPPP}
\multicolumn{23}{c}{Two-block $X$ measurement} \\
Total ancilla size & \multicolumn{2}{c}{$X_0 \otimes X_1$} & & & & & & & & & & & & & & & & & & & \\
%\hline
78 & \bluecell{2} & \orangecell{4} &  &  &  &  &  &  &  &  &  &  &  &  &  &  &  &  &  &  &  &  \\
\hline\hline
\multicolumn{23}{c}{Two-block $Z$ measurement} \\
Total ancilla size & \multicolumn{2}{c}{$Z_2 \otimes Z_3$} & & & & & & & & & & & & & & & & & & & \\
%\hline
 54 & \bluecell{4} & \orangecell{4} &  &  &  &  &  &  &  &  &  &  &  &  &  &  &  &  &  &  &  &  \\
\hline
\end{tabular}
\caption{
Code distances of deformed codes for the [[54,6,4]] BB code. 
}
\label{table:54}
\end{table*}

\begin{table*}[t]
\centering
\small
\begin{tabular}{cPPPPPPPPPPPPPPPPPPPPPPPPPP}
\hline\hline
\multicolumn{25}{c}{Single-block $X$ measurement} \\
Total ancilla size & \multicolumn{2}{c}{$X_0$} & \multicolumn{2}{c}{$X_1$} & \multicolumn{2}{c}{$X_2$} & \multicolumn{2}{c}{$X_3$} & \multicolumn{2}{c}{$X_4$} & \multicolumn{2}{c}{$X_5$}& \multicolumn{2}{c}{$X_6$}& \multicolumn{2}{c}{$X_7$} & \multicolumn{2}{c}{$X_0X_1$} & \multicolumn{2}{c}{$X_0X_2$} & \multicolumn{2}{c}{$X_0X_3$} & \multicolumn{2}{c}{$X_0X_4$}  \\
28 & \bluecell{2} & \orangecell{3} & \bluecell{2} & \orangecell{4} & \bluecell{2} & \orangecell{2} & \bluecell{2} & \orangecell{2} & \bluecell{2} & \orangecell{3} & \bluecell{2} & \orangecell{4} & \bluecell{2} & \orangecell{4} & \bluecell{3} & \orangecell{4} & \bluecell{2} & \orangecell{3} & \bluecell{2} & \orangecell{4} & \bluecell{5} & \orangecell{5} & \bluecell{3} & \orangecell{4} \\
46 & \bluecell{2} & \orangecell{4} & \bluecell{2} & \orangecell{4} & \bluecell{2} & \orangecell{3} & \bluecell{2} & \orangecell{3} & \bluecell{2} & \orangecell{4} & \bluecell{2} & \orangecell{5} & \bluecell{2} & \orangecell{5} & \bluecell{5} & \orangecell{5} & \bluecell{2} & \orangecell{4} & \bluecell{2} & \orangecell{4} & \bluecell{5} & \orangecell{5} & \bluecell{2} & \orangecell{4} \\
64 & \bluecell{3} & \orangecell{5} & \bluecell{4} & \orangecell{6} & \bluecell{2} & \orangecell{6} & \bluecell{2} & \orangecell{6} & \bluecell{2} & \orangecell{6} & \bluecell{3} & \orangecell{6} & \bluecell{3} & \orangecell{6} & \bluecell{6} & \orangecell{6} & \bluecell{2} & \orangecell{4} & \bluecell{2} & \orangecell{6} & \bluecell{7} & \orangecell{7} & \bluecell{2} & \orangecell{6} \\
82 & \bluecell{6} & \orangecell{7} & \bluecell{4} & \orangecell{6} & \bluecell{3} & \orangecell{6} & \bluecell{2} & \orangecell{7} & \bluecell{2} & \orangecell{6} & \bluecell{4} & \orangecell{7} & \bluecell{3} & \orangecell{6} & \bluecell{7} & \orangecell{7} & \bluecell{3} & \orangecell{6} & \bluecell{2} & \orangecell{6} &  &  & \bluecell{3} & \orangecell{6} \\
100 &  &  & \bluecell{6} & \orangecell{7} & \bluecell{3} & \orangecell{6} &  &  & \bluecell{3} & \orangecell{6} &  &  & \bluecell{3} & \orangecell{7} &  &  & \bluecell{2} & \orangecell{7} & \bluecell{3} & \orangecell{6} &  &  & \bluecell{4} & \orangecell{7} \\
118 &  &  &  &  & \bluecell{3} & \orangecell{6} &  &  & \bluecell{2} & \orangecell{7} &  &  &  &  &  &  &  &  & \bluecell{2} & \orangecell{6} &  &  &  &  \\
136 &  &  &  &  & \bluecell{3} & \orangecell{7} &  &  &  &  &  &  &  &  &  &  &  &  & \bluecell{3} & \orangecell{7} &  &  &  &  \\

 & \multicolumn{2}{c}{$X_0X_5$} & \multicolumn{2}{c}{$X_0X_6$} & \multicolumn{2}{c}{$X_0X_7$} & \multicolumn{2}{c}{$X_1X_2$} & \multicolumn{2}{c}{$X_1X_3$} & \multicolumn{2}{c}{$X_1X_4$}& \multicolumn{2}{c}{$X_1X_5$}& \multicolumn{2}{c}{$X_1X_6$} & \multicolumn{2}{c}{$X_1X_7$} & \multicolumn{2}{c}{$X_2X_3$} & \multicolumn{2}{c}{$X_2X_4$} & \multicolumn{2}{c}{$X_2X_5$}  \\
28 & \bluecell{3} & \orangecell{3} & \bluecell{2} & \orangecell{4} & \bluecell{2} & \orangecell{4} & \bluecell{2} & \orangecell{2} & \bluecell{2} & \orangecell{2} & \bluecell{2} & \orangecell{3} & \bluecell{2} & \orangecell{4} & \bluecell{2} & \orangecell{4} & \bluecell{3} & \orangecell{4} & \bluecell{2} & \orangecell{4} & \bluecell{2} & \orangecell{5} & \bluecell{2} & \orangecell{2} \\
46 & \bluecell{4} & \orangecell{5} & \bluecell{2} & \orangecell{4} & \bluecell{3} & \orangecell{4} & \bluecell{2} & \orangecell{3} & \bluecell{2} & \orangecell{3} & \bluecell{2} & \orangecell{4} & \bluecell{2} & \orangecell{5} & \bluecell{3} & \orangecell{4} & \bluecell{2} & \orangecell{5} & \bluecell{2} & \orangecell{4} & \bluecell{2} & \orangecell{5} & \bluecell{2} & \orangecell{3} \\
64 & \bluecell{3} & \orangecell{5} & \bluecell{2} & \orangecell{6} & \bluecell{3} & \orangecell{6} & \bluecell{3} & \orangecell{6} & \bluecell{2} & \orangecell{4} & \bluecell{2} & \orangecell{6} & \bluecell{2} & \orangecell{6} & \bluecell{4} & \orangecell{6} & \bluecell{2} & \orangecell{6} & \bluecell{2} & \orangecell{6} & \bluecell{2} & \orangecell{7} & \bluecell{3} & \orangecell{6} \\
82 & \bluecell{3} & \orangecell{7} & \bluecell{2} & \orangecell{6} & \bluecell{5} & \orangecell{6} & \bluecell{5} & \orangecell{6} & \bluecell{2} & \orangecell{5} & \bluecell{2} & \orangecell{6} & \bluecell{4} & \orangecell{7} & \bluecell{3} & \orangecell{6} & \bluecell{2} & \orangecell{7} & \bluecell{2} & \orangecell{7} &  &  & \bluecell{2} & \orangecell{6} \\
100 &  &  & \bluecell{2} & \orangecell{6} & \bluecell{5} & \orangecell{6} & \bluecell{6} & \orangecell{7} & \bluecell{2} & \orangecell{7} & \bluecell{2} & \orangecell{7} &  &  & \bluecell{3} & \orangecell{7} &  &  &  &  &  &  & \bluecell{2} & \orangecell{6} \\
118 &  &  & \bluecell{2} & \orangecell{7} & \bluecell{7} & \orangecell{7} &  &  &  &  &  &  &  &  &  &  &  &  &  &  &  &  & \bluecell{3} & \orangecell{7} \\

 & \multicolumn{2}{c}{$X_2X_6$} & \multicolumn{2}{c}{$X_2X_7$} & \multicolumn{2}{c}{$X_3X_4$} & \multicolumn{2}{c}{$X_3X_5$} & \multicolumn{2}{c}{$X_3X_6$} & \multicolumn{2}{c}{$X_3X_7$}& \multicolumn{2}{c}{$X_4X_5$}& \multicolumn{2}{c}{$X_4X_6$} & \multicolumn{2}{c}{$X_4X_7$} & \multicolumn{2}{c}{$X_5X_6$} & \multicolumn{2}{c}{$X_5X_7$} & \multicolumn{2}{c}{$X_6X_7$}  \\
28 & \bluecell{2} & \orangecell{4} & \bluecell{2} & \orangecell{2} & \bluecell{6} & \orangecell{6} & \bluecell{3} & \orangecell{4} & \bluecell{2} & \orangecell{4} & \bluecell{4} & \orangecell{4} & \bluecell{3} & \orangecell{4} & \bluecell{2} & \orangecell{3} & \bluecell{3} & \orangecell{4} & \bluecell{2} & \orangecell{2} & \bluecell{2} & \orangecell{4} & \bluecell{2} & \orangecell{4} \\
46 & \bluecell{2} & \orangecell{4} & \bluecell{2} & \orangecell{3} & \bluecell{6} & \orangecell{6} & \bluecell{2} & \orangecell{4} & \bluecell{2} & \orangecell{4} & \bluecell{3} & \orangecell{5} & \bluecell{3} & \orangecell{4} & \bluecell{2} & \orangecell{3} & \bluecell{3} & \orangecell{4} & \bluecell{2} & \orangecell{3} & \bluecell{3} & \orangecell{4} & \bluecell{2} & \orangecell{6} \\
64 & \bluecell{3} & \orangecell{6} & \bluecell{2} & \orangecell{5} & \bluecell{7} & \orangecell{7} & \bluecell{3} & \orangecell{6} & \bluecell{2} & \orangecell{6} & \bluecell{3} & \orangecell{6} & \bluecell{2} & \orangecell{6} & \bluecell{2} & \orangecell{4} & \bluecell{5} & \orangecell{6} & \bluecell{2} & \orangecell{5} & \bluecell{4} & \orangecell{6} & \bluecell{2} & \orangecell{6} \\
82 & \bluecell{5} & \orangecell{7} & \bluecell{2} & \orangecell{6} &  &  & \bluecell{3} & \orangecell{6} & \bluecell{2} & \orangecell{6} & \bluecell{5} & \orangecell{7} & \bluecell{3} & \orangecell{6} & \bluecell{2} & \orangecell{5} & \bluecell{5} & \orangecell{7} & \bluecell{2} & \orangecell{5} & \bluecell{5} & \orangecell{6} & \bluecell{2} & \orangecell{7} \\
100 &  &  & \bluecell{3} & \orangecell{7} &  &  & \bluecell{5} & \orangecell{6} & \bluecell{3} & \orangecell{6} &  &  & \bluecell{5} & \orangecell{7} & \bluecell{2} & \orangecell{7} &  &  & \bluecell{2} & \orangecell{7} & \bluecell{4} & \orangecell{6} &  &  \\
118 &  &  &  &  &  &  & \bluecell{5} & \orangecell{6} & \bluecell{3} & \orangecell{6} &  &  &  &  &  &  &  &  &  &  & \bluecell{3} & \orangecell{7} &  &  \\
136 &  &  &  &  &  &  & \bluecell{7} & \orangecell{7} & \bluecell{3} & \orangecell{7} &  &  &  &  &  &  &  &  &  &  &  &  &  &  \\

\hline\hline
\multicolumn{23}{c}{Single-block $Z$ measurement} \\
Total ancilla size & \multicolumn{2}{c}{$Z_0$} & \multicolumn{2}{c}{$Z_1$} & \multicolumn{2}{c}{$Z_2$} & \multicolumn{2}{c}{$Z_3$} & \multicolumn{2}{c}{$Z_4$} & \multicolumn{2}{c}{$Z_5$}& \multicolumn{2}{c}{$Z_6$}& \multicolumn{2}{c}{$Z_7$} & \multicolumn{2}{c}{$Z_0Z_1$} & \multicolumn{2}{c}{$Z_0Z_2$} & \multicolumn{2}{c}{$Z_0Z_3$} & \multicolumn{2}{c}{$Z_0Z_4$}  \\
30 & \bluecell{2} & \orangecell{4} & \bluecell{2} & \orangecell{3} & \bluecell{2} & \orangecell{5} & \bluecell{2} & \orangecell{5} & \bluecell{3} & \orangecell{6} & \bluecell{2} & \orangecell{5} & \bluecell{2} & \orangecell{5} & \bluecell{2} & \orangecell{4} & \bluecell{2} & \orangecell{4} & \bluecell{2} & \orangecell{5} & \bluecell{2} & \orangecell{4} & \bluecell{2} & \orangecell{3} \\
48 & \bluecell{2} & \orangecell{5} & \bluecell{2} & \orangecell{5} & \bluecell{2} & \orangecell{6} & \bluecell{2} & \orangecell{6} & \bluecell{4} & \orangecell{7} & \bluecell{5} & \orangecell{6} & \bluecell{4} & \orangecell{6} & \bluecell{2} & \orangecell{5} & \bluecell{5} & \orangecell{5} & \bluecell{2} & \orangecell{6} & \bluecell{2} & \orangecell{5} & \bluecell{2} & \orangecell{4} \\
66 & \bluecell{2} & \orangecell{6} & \bluecell{2} & \orangecell{5} & \bluecell{3} & \orangecell{6} & \bluecell{2} & \orangecell{7} &  &  & \bluecell{6} & \orangecell{6} & \bluecell{4} & \orangecell{6} & \bluecell{2} & \orangecell{6} & \bluecell{4} & \orangecell{6} & \bluecell{2} & \orangecell{6} & \bluecell{2} & \orangecell{6} & \bluecell{2} & \orangecell{5} \\
84 & \bluecell{3} & \orangecell{6} & \bluecell{2} & \orangecell{6} & \bluecell{2} & \orangecell{7} &  &  &  &  & \bluecell{7} & \orangecell{7} & \bluecell{5} & \orangecell{6} & \bluecell{2} & \orangecell{6} & \bluecell{2} & \orangecell{6} & \bluecell{2} & \orangecell{6} & \bluecell{3} & \orangecell{7} & \bluecell{2} & \orangecell{6} \\
102 & \bluecell{4} & \orangecell{7} & \bluecell{3} & \orangecell{6} &  &  &  &  &  &  &  &  & \bluecell{6} & \orangecell{7} & \bluecell{2} & \orangecell{6} & \bluecell{7} & \orangecell{7} & \bluecell{2} & \orangecell{6} &  &  & \bluecell{3} & \orangecell{6} \\
120 &  &  & \bluecell{2} & \orangecell{7} &  &  &  &  &  &  &  &  &  &  & \bluecell{2} & \orangecell{7} &  &  & \bluecell{3} & \orangecell{7} &  &  & \bluecell{4} & \orangecell{6} \\
138 &  &  &  &  &  &  &  &  &  &  &  &  &  &  &  &  &  &  &  &  &  &  & \bluecell{4} & \orangecell{7} \\
 & \multicolumn{2}{c}{$Z_0Z_5$} & \multicolumn{2}{c}{$Z_0Z_6$} & \multicolumn{2}{c}{$Z_0Z_7$} & \multicolumn{2}{c}{$Z_1Z_2$} & \multicolumn{2}{c}{$Z_1Z_3$} & \multicolumn{2}{c}{$Z_1Z_4$}& \multicolumn{2}{c}{$Z_1Z_5$}& \multicolumn{2}{c}{$Z_1Z_6$} & \multicolumn{2}{c}{$Z_1Z_7$} & \multicolumn{2}{c}{$Z_2Z_3$} & \multicolumn{2}{c}{$Z_2Z_4$} & \multicolumn{2}{c}{$Z_2Z_5$}  \\

30 & \bluecell{3} & \orangecell{4} & \bluecell{4} & \orangecell{4} & \bluecell{2} & \orangecell{4} & \bluecell{2} & \orangecell{3} & \bluecell{2} & \orangecell{4} & \bluecell{3} & \orangecell{4} & \bluecell{2} & \orangecell{2} & \bluecell{2} & \orangecell{4} & \bluecell{2} & \orangecell{4} & \bluecell{2} & \orangecell{4} & \bluecell{2} & \orangecell{5} & \bluecell{2} & \orangecell{5} \\
48 & \bluecell{2} & \orangecell{5} & \bluecell{4} & \orangecell{6} & \bluecell{4} & \orangecell{4} & \bluecell{2} & \orangecell{4} & \bluecell{2} & \orangecell{6} & \bluecell{2} & \orangecell{4} & \bluecell{2} & \orangecell{3} & \bluecell{2} & \orangecell{5} & \bluecell{2} & \orangecell{5} & \bluecell{2} & \orangecell{5} & \bluecell{2} & \orangecell{5} & \bluecell{2} & \orangecell{6} \\
66 & \bluecell{2} & \orangecell{6} & \bluecell{4} & \orangecell{6} & \bluecell{2} & \orangecell{6} & \bluecell{3} & \orangecell{5} & \bluecell{4} & \orangecell{6} & \bluecell{2} & \orangecell{5} & \bluecell{2} & \orangecell{5} & \bluecell{2} & \orangecell{7} & \bluecell{5} & \orangecell{6} & \bluecell{4} & \orangecell{6} & \bluecell{3} & \orangecell{6} & \bluecell{3} & \orangecell{6} \\
84 & \bluecell{3} & \orangecell{7} & \bluecell{4} & \orangecell{6} & \bluecell{3} & \orangecell{7} & \bluecell{4} & \orangecell{6} & \bluecell{5} & \orangecell{7} & \bluecell{3} & \orangecell{6} & \bluecell{3} & \orangecell{6} &  &  & \bluecell{6} & \orangecell{7} & \bluecell{4} & \orangecell{6} & \bluecell{3} & \orangecell{7} & \bluecell{4} & \orangecell{6} \\
102 &  &  & \bluecell{4} & \orangecell{7} &  &  & \bluecell{5} & \orangecell{7} &  &  & \bluecell{4} & \orangecell{6} & \bluecell{2} & \orangecell{7} &  &  &  &  & \bluecell{5} & \orangecell{6} &  &  & \bluecell{2} & \orangecell{6} \\
120 &  &  &  &  &  &  &  &  &  &  & \bluecell{2} & \orangecell{7} &  &  &  &  &  &  & \bluecell{6} & \orangecell{7} &  &  & \bluecell{2} & \orangecell{6} \\
138 &  &  &  &  &  &  &  &  &  &  &  &  &  &  &  &  &  &  &  &  &  &  & \bluecell{2} & \orangecell{7} \\
 & \multicolumn{2}{c}{$Z_2Z_6$} & \multicolumn{2}{c}{$Z_2Z_7$} & \multicolumn{2}{c}{$Z_3Z_4$} & \multicolumn{2}{c}{$Z_3Z_5$} & \multicolumn{2}{c}{$Z_3Z_6$} & \multicolumn{2}{c}{$Z_3Z_7$}& \multicolumn{2}{c}{$Z_4Z_5$}& \multicolumn{2}{c}{$Z_4Z_6$} & \multicolumn{2}{c}{$Z_4Z_7$} & \multicolumn{2}{c}{$Z_5Z_6$} & \multicolumn{2}{c}{$Z_5Z_7$} & \multicolumn{2}{c}{$Z_6Z_7$}  \\
30 & \bluecell{2} & \orangecell{3} & \bluecell{2} & \orangecell{3} & \bluecell{2} & \orangecell{4} & \bluecell{2} & \orangecell{5} & \bluecell{2} & \orangecell{5} & \bluecell{2} & \orangecell{5} & \bluecell{2} & \orangecell{5} & \bluecell{2} & \orangecell{4} & \bluecell{2} & \orangecell{5} & \bluecell{3} & \orangecell{5} & \bluecell{2} & \orangecell{6} & \bluecell{4} & \orangecell{5} \\
48 & \bluecell{2} & \orangecell{4} & \bluecell{2} & \orangecell{4} & \bluecell{2} & \orangecell{5} & \bluecell{2} & \orangecell{6} & \bluecell{2} & \orangecell{5} & \bluecell{3} & \orangecell{6} & \bluecell{2} & \orangecell{6} & \bluecell{3} & \orangecell{5} & \bluecell{2} & \orangecell{5} & \bluecell{3} & \orangecell{6} & \bluecell{2} & \orangecell{7} & \bluecell{2} & \orangecell{6} \\
66 & \bluecell{2} & \orangecell{5} & \bluecell{3} & \orangecell{6} & \bluecell{3} & \orangecell{6} & \bluecell{2} & \orangecell{6} & \bluecell{2} & \orangecell{6} & \bluecell{5} & \orangecell{7} & \bluecell{3} & \orangecell{6} & \bluecell{2} & \orangecell{5} & \bluecell{2} & \orangecell{6} & \bluecell{2} & \orangecell{6} &  &  & \bluecell{2} & \orangecell{6} \\
84 & \bluecell{2} & \orangecell{7} & \bluecell{2} & \orangecell{6} & \bluecell{2} & \orangecell{7} & \bluecell{4} & \orangecell{7} & \bluecell{2} & \orangecell{6} &  &  & \bluecell{5} & \orangecell{7} & \bluecell{2} & \orangecell{6} & \bluecell{3} & \orangecell{7} & \bluecell{2} & \orangecell{7} &  &  & \bluecell{3} & \orangecell{7} \\
102 &  &  & \bluecell{2} & \orangecell{6} &  &  &  &  & \bluecell{2} & \orangecell{6} &  &  &  &  & \bluecell{3} & \orangecell{7} &  &  &  &  &  &  &  &  \\
120 &  &  & \bluecell{2} & \orangecell{7} &  &  &  &  & \bluecell{2} & \orangecell{7} &  &  &  &  &  &  &  &  &  &  &  &  &  &  \\

\hline\hline
\end{tabular}

\begin{tabular}{cPPPPPPPPPPPPPPPPPPPPPPPPPP}
\multicolumn{25}{c}{Two-block $X$ measurement} \\
Total ancilla size & \multicolumn{2}{c}{$X_0 \otimes X_1$} & & & & & & & & & & & & & & & & & & & & & \\
% \hline
162 & \bluecell{3} & \orangecell{7} &            &              &            &              &            &              &            &              &            &              &            &              &            &              &            &              &            &              &  & &          &              \\
\hline\hline
\multicolumn{25}{c}{Two-block $Z$ measurement} \\
Total ancilla size & \multicolumn{2}{c}{$Z_2 \otimes Z_3$} & & & & & & & & & & & & & & & & & & & & & \\
% \hline
 144 & \bluecell{3} & \orangecell{7} &            &              &            &              &            &              &            &              &            &              &            &              &            &              &            &              &            &              &            &              \\
\hline
\end{tabular}
\caption{
Code distances of deformed codes for the [[162,8,7]] BB code. 
}
\label{table:162}
\end{table*}

\subsection{Ancilla size for fault-tolerant logical measurements}
\label{sec:NANO}

In this section, we present additional numerical simulations across a broader range of codes and logical operators to validate that the code distance is preserved in our protocol.

The results are summarized in Fig.~\ref{fig:NANO}. The codes $[[144,4,8]]$ and $[[245,6,5]]$ are directional $N^2ESEN^2$ and $N^2E^3N^2$ codes adapted to open boundaries, respectively, constructed from $7\times7$ and $12\times7$ copies of bulk tiles~\cite{geher2025directional}. As a systematic construction and search for the planar-directional category have not yet been reported, we performed a preliminary construction and search ourselves. Since an exhaustive search for optimal codes is beyond the scope of this study, our efforts were limited; consequently, the parameters for the planar directional codes remain subject to further optimization. The code $[[88,6,6]]$ is taken from Ref.~\cite{Liang2025}. The codes $[[288,8,12]]$, $[[288,18,13]]$, $[[288,18,14]]$, $[[512,18,19]]$, and $[[512,18,\leq 21]]$ are taken from Ref.~\cite{Steffan2025}.

For each code, we randomly sampled ten nontrivial $X$-type logical operators from the logical operator space, which contains $2^k-1$ elements. For each sampled operator, we constructed the corresponding deformed code implementing a fault-tolerant logical measurement and determined the minimum ancilla-system size $N_A$ required to preserve the original code distance. 

For codes with distance $d > 19$, exact distance evaluation using integer linear programming~\cite{landahl2011fault} becomes computationally prohibitive due to its exponential runtime scaling. Accordingly, in our protocol we estimate the code distance using the heuristic integer linear programming algorithm~\cite{gurobi}, which provides an upper bound on the true distance. While this limits the precision of distance verification at large $d$, it remains sufficient to confirm distance preservation within the numerical accuracy accessible to current methods.

Across all evaluated instances, we find that our protocol successfully realizes the targeted fault-tolerant logical measurements while preserving the original code distance. Regarding qubit overhead, we observe a general trend of increasing ancilla-system size with increasing code distance. Nevertheless, even for distances as high as $21$, the ancilla system remains comparable in size to the original code block. In particular, the maximum observed overhead ratio is $N_A/N_O = 1.582$, demonstrating that the protocol remains qubit-efficient in the distance regime relevant for early-stage fault-tolerant quantum computation.

\begin{figure}[htbp]
\centering
\includegraphics[width=0.8\linewidth]{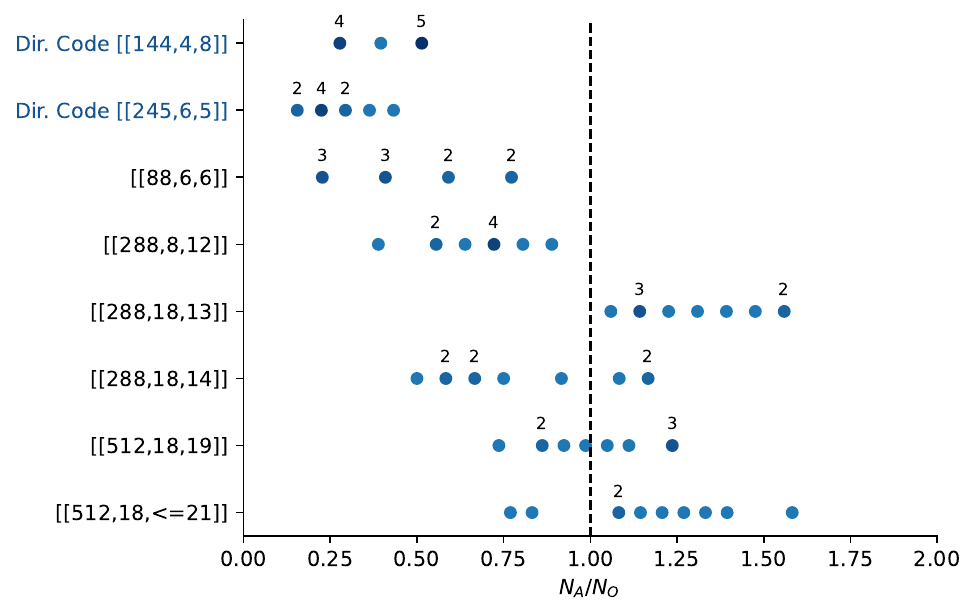}
\caption{Relative ancilla system size for fault-tolerant logical measurements. $N_O$ denotes the number of physical qubits in the original code block, and $N_A$ denotes the number of physical qubits in the ancilla system. For each code, we sampled $10$ random logical operators and evaluated the minimum $N_A$ required to preserve the original code distance. The multiplicity of logical operators requiring specific ancilla sizes is illustrated in the figure. Notably, the top two codes (highlighted in blue) correspond to open-boundary directional codes. The code distance $\leq 21$ is obtained via the heuristic integer linear programming algorithm; therefore, the result is an upper bound estimate.
}
\label{fig:NANO}
\end{figure}

\section{Resource Analysis}

In this section, we provide a theoretical analysis of the space and time resources required by our protocol. In particular, we estimate the overhead incurred by incorporating surface-code patches to complete the logical gate set. Our analysis focuses on the implementation of a single layer of logical operations drawn from the universal gate set $\{H, S, T, \mathrm{CNOT}\}$.

\begin{figure}[htbp]
\centering
\includegraphics[width=0.4\linewidth]{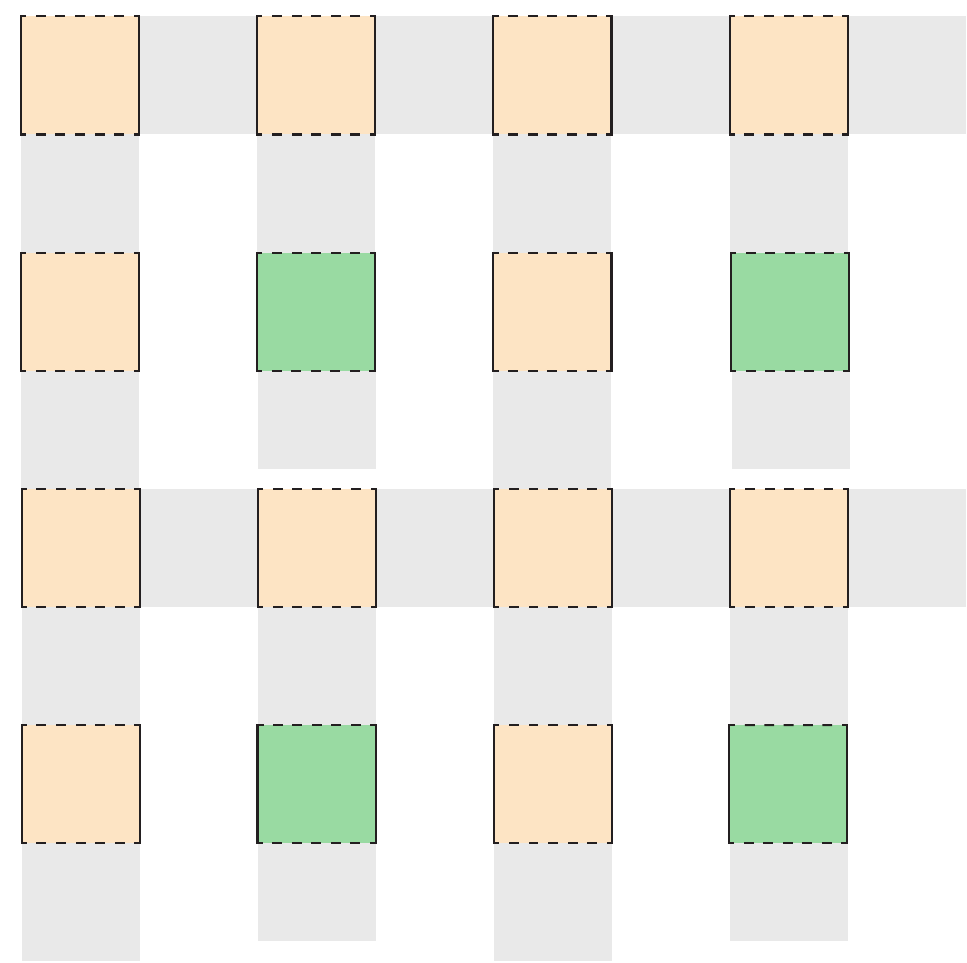}
\caption{
Array of code blocks. Orange squares denote planar-BB-code blocks, while the green square indicates a surface‑code block. Gray squares represent ancilla regions, which are used to perform code surgery. 
}
\label{fig:block_array}
\end{figure}

\textbf{Density of surface-code blocks.} The time overhead depends on the density of surface-code blocks. We assume that each surface-code block services $m$ high-rate code blocks. We refer to the group of $m$ high-rate blocks sharing a surface-code block as a \emph{block cluster} (noting that a surface-code block may be temporarily borrowed by neighboring clusters in practical implementations). A representative block arrangement for the case $m = 3$ is illustrated in Fig.~\ref{fig:block_array}.

\begin{figure}[htbp]
\centering
\includegraphics[width=0.4\linewidth]{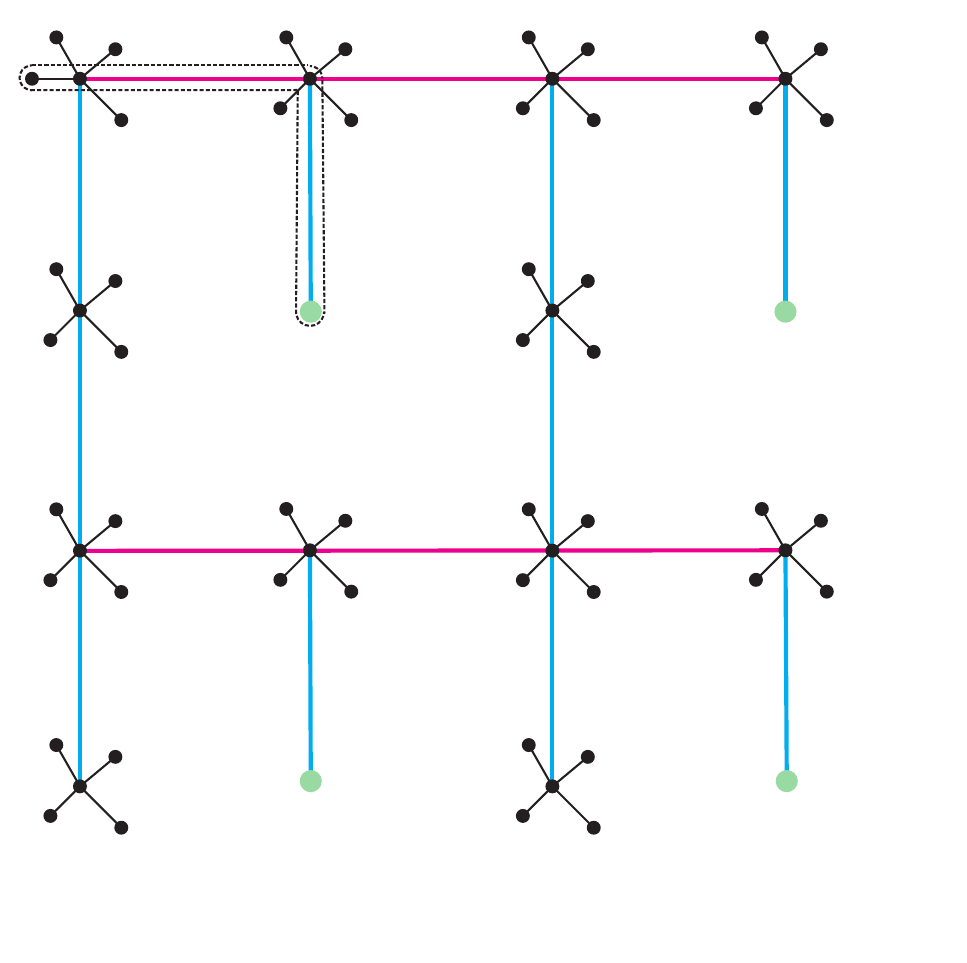}
\caption{
Network of logical qubits. Each black circle represents a logical qubit encoded in a planar-BB-code block, and each green circle represents a surface-code logical qubit. Red and blue edges depict $XX$ and $ZZ$ joint measurements, respectively; black edges signify that both types of joint measurements can be performed between the connected qubits. The dashed box highlights the routing path between a data logical qubit and a surface-code logical qubit.
}
\label{fig:logical_network}
\end{figure}

\textbf{Overview of the setup.} To construct the logical-qubit network, we designate one logical qubit within each block as the central logical qubit, which is coupled to all other logical qubits in the same block. While our numerical benchmarks demonstrate that arbitrary intra-block $X$- or $Z$-type joint measurements are achievable between any pair of logical qubits, we restrict our analysis to a star-topology configuration for simplicity. In this setup, we only utilize intra-block joint measurements involving the central logical qubit during logical-circuit compilation. Furthermore, for simplicity, we assume that each central qubit is coupled to its counterparts (or to surface-code logical qubits) in nearest-neighboring blocks, as illustrated in Fig.~\ref{fig:logical_network}. Within this architecture, the central logical qubits function as mediating ancillas, while the remaining qubits are reserved for data storage.

For the purpose of theoretical analysis, we assume that the quantum computer operates as follows: at each time step, at most one logical gate is applied within each block cluster.

\begin{figure}[htbp]
\centering
\includegraphics[width=0.4\linewidth]{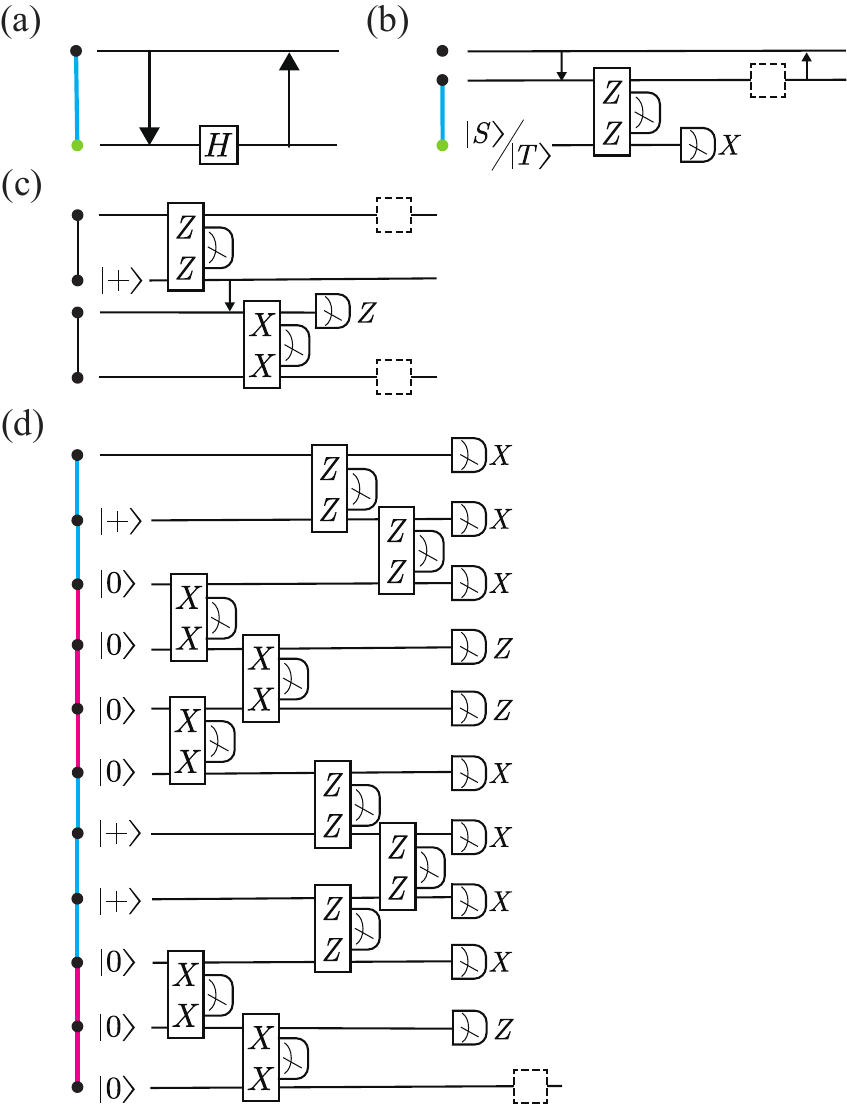}
\caption{
(a) Circuit for the $H$ gate. (b) Circuit for the $S$ and $T$ gates. (c) Circuit for the CNOT gate. Arrows denote state transfers between logical qubits, while dashed boxes indicate feedback gates. (d) State transfer circuit. Interactions between central logical qubits are restricted to either $XX$ or $ZZ$ joint measurements. The routing path may alternate between these two types of couplings.
}
\label{fig:logical_gates}
\end{figure}

\textbf{Single-qubit gates.} We utilize a surface-code block to implement the $H$, $S$, and $T$ gates, with the corresponding logical circuits shown in Figs.~\ref{fig:logical_gates}(a,b).

The $H$ gate is implemented by transferring the state of the target logical qubit to the surface-code block, applying the logical Hadamard gate~\cite{Fowler2012,Horsman2012}, and returning the state to the logical qubit. This state transfer is mediated by a chain of central logical qubits, as illustrated in Figs.~\ref{fig:logical_network} and \ref{fig:logical_gates}(d).

In terms of time overhead, the surface-code Hadamard gate and logical measurements each require $O(d)$ parity-check cycles. While the exact cycle count depends on hardware-specific parameters (such as the relative rates of qubit and measurement errors), the total time complexity for implementing an $H$ gate scales asymptotically as $O(d)$.

For the $S$ and $T$ gates, the required magic state is first prepared on a surface-code block~\cite{Bravyi2012,Li2014AMS,2024arXiv240303991I,Gidney2024,Chen2025}. We then route the state of the target logical qubit to the central logical qubit adjacent to the surface-code block. Following the gate execution on the central logical qubit, the state is transferred back to the target logical qubit. Since magic-state preparation (assuming magic state cultivation) requires $O(d)$ parity-check cycles, the overall time overhead for the $S$ and $T$ gates also scales as $O(d)$.

\textbf{CNOT gate.} A logical CNOT gate is realized through logical measurements and requires an ancilla logical qubit. When the control and target logical qubits reside in the same code block, the central logical qubit is assigned as the ancilla. For inter-block CNOT gates, the operation is initiated by employing the central logical qubit of the control block as the ancilla; the state is then transferred to the central logical qubit of the target block to complete the operation [see Fig.~\ref{fig:logical_gates}(c).]. Consequently, the total time overhead for implementing a CNOT gate scales as $O(d)$.

\textbf{Qubit Overhead.} While a theoretical upper bound for the ancilla system size required for logical measurements has yet to be established, our numerical simulations across various codes and logical operators indicate that the required ancilla size is comparable to the code block size. Based on this observation, we estimate the qubit overhead using the following assumptions:

\begin{enumerate}
    \item Each high-rate code block comprises approximately $2n$ physical qubits (consisting of $n$ data qubits and $n$ ancilla qubits for parity checks).
    \item Each ancilla system used for logical measurement on high-rate code blocks consists of approximately $2\chi n$ physical qubits, where $\chi \approx 1$ is a scaling factor based on our numerical findings.
    \item Each surface code block contains approximately $2d^2$ physical qubits (assuming a rotated surface code layout). An additional $2d^2$ physical qubits are required to facilitate logical Hadamard gates via lattice-surgery-type operations.
    \item A coupling ancilla system is required between each surface-code block and a high-rate code block. Following the CKBB protocol~\cite{Cohen2022,Xu2024}, the number of physical qubits required for this region is approximately $4wd$, where $w$ is the weight of the corresponding logical operator. For simplicity, we assume $w = d$.
\end{enumerate}

Consequently, for each cluster containing $m$ high-rate blocks, the total number of physical qubits is 
\begin{equation}
    N_P = \underbrace{2nm}_{\text{high-rate blocks}} + \underbrace{2n\chi(m+1)}_{\text{measurement ancillas}} + \underbrace{2d^2 + 2d^2 + 4d^2}_{\text{surface-code resources}} = 2n[m + (m+1)\chi] + 8d^2.
\end{equation}
This formula accounts for $m$ high-rate blocks ($2nm$), $m+1$ measurement ancilla systems [$2n\chi(m+1)$], and the surface-code component ($2d^2 \text{ data} + 2d^2 \text{ Hadamard ancilla} + 4d^2 \text{ coupling region}$). Given that each cluster supports $N_L = m(k - 1)$ data logical qubits, the resulting qubit overhead $N_P/N_L$ is given in Table~\ref{tab:overhead}.

\begin{table}[htbp]
    \centering
    \begin{tabular}{@{}c@{\hspace{8pt}} c@{\hspace{8pt}} c@{\hspace{8pt}} c@{\hspace{8pt}} c@{\hspace{8pt}} c@{\hspace{8pt}} c@{\hspace{8pt}} c@{\hspace{8pt}} c@{\hspace{8pt}}
                    @{\hspace{8pt}}|@{\hspace{8pt}} c@{\hspace{8pt}}}
        \hline\hline
        $m=$ & 4 & 9 & 25 & 36 & 49 & 64 & 81 & 100 & \text{Surface code} \\
        \hline
        Dir. Code $[[144,4,8]]$      & 200.333& 169.778& 154.133& 151.444& 149.823& 148.771& 148.049& 147.533
& 256 \\
        Dir. Code $[[245,6,5]]$      & 161& 149.556& 143.696& 142.689& 142.082& 141.688& 141.417& 141.224
& 100 \\
        $[[88,6,6]]$                 & 83.6& 71.822& 65.792& 64.756& 64.131& 63.725& 63.447& 63.248
& 144 \\
        $[[288,8,12]]$               & 214.857& 181.841& 164.937& 162.032& 160.28& 159.143& 158.363& 157.806
& 576 \\
        $[[288,18,13]]$              & 119.794& 101.412& 92& 90.382& 89.407& 88.774& 88.34& 88.029
& 676 \\
        $[[288,18,14]]$              & 106.353& 88.052& 78.682& 77.072& 76.101& 75.471& 75.038& 74.729
& 784 \\
        $[[512,18,19]]$              & 195.794& 161.856& 144.48& 141.493& 139.693& 138.524& 137.723& 137.149
& 1444 \\
        $[[512,18,\leq 21]]$         & 231.235& 189.176& 167.642& 163.941& 161.709& 160.261& 159.268& 158.558
& 1764 \\
        \hline
    \end{tabular}
    \caption{Qubit overhead $N_P/N_L$. Here, $N_P$ denotes the total number of physical qubits, and $N_L$ denotes the number of supported data logical qubits within the same cluster. Each column corresponds to a block cluster of size $m$, defined as the number of high-rate planar code blocks sharing a single surface-code block. 
    For an $[[n,k,d]]$ code, the number of logical qubits per cluster is $N_L = m(k - 1)$, while the total number of physical qubits is estimated as $N_P = 2n[m + (m+1)\chi] + 8d^2$. The parameter $\chi$ is determined from numerical simulations as the maximum ancilla-system overhead, $\chi=\max \{N_A/N_O\}$, observed over random logical operators (Fig.~\ref{fig:NANO}). Results are shown for various cluster sizes $m$ and are compared to the surface-code architecture, whose effective qubit overhead is taken to be $4d^2$ physical qubits per logical qubit (final column).}
    \label{tab:overhead}
\end{table}

In contrast, in a standard surface-code architecture, each logical qubit is encoded using about $2d^2$ physical qubits. While individual logical measurements can be performed with minimal ancilla overhead (e.g., a one-dimensional chain of qubits), additional ancilla blocks are necessary between data blocks to facilitate $H$ and CNOT gates~\cite{Litinski2019}. Therefore, the effective overhead for a pure surface-code architecture is approximately $4d^2$ physical qubits per logical qubit.

\textbf{Time Overhead.} In the following time overhead analysis, we assume that CNOT gates are restricted to local interactions---acting either within a cluster or between nearest-neighboring clusters---reflecting the constraints of an underlying two-dimensional physical qubit network. Long-range gates are handled via logical-level circuit compilation and are outside the scope of this analysis. Under these constraints, CNOT gates can be executed in parallel provided they act on different clusters.

We estimate the time overhead by considering a single circuit layer, defined as a set of operations where at most one logical gate (either a single-qubit gate or a local CNOT) is applied to each data logical qubit. To complete such a layer in our architecture, the required time scales as $O(mkd)$. While individual logical gates require $O(d)$ time steps, the architecture introduces a serialization penalty: since the shared resources within a cluster allow only one logical gate to be processed at a time for the $m(k-1)$ data qubits, the effective time overhead for a full layer scales with the number of data logical qubits in the cluster, resulting in $O(mkd)$.

This locality constraint is similarly present in pure surface-code architectures on two-dimensional networks. If we consider only CNOT gates between nearest-neighboring data logical qubits in a surface-code array, the time overhead is simply $O(d)$. However, this does not imply that our approach suffers from an additional time overhead factor of $O(mk)$. Rather, the two approaches result in logical qubit networks with fundamentally different topologies. For the surface code, the logical network is restricted to a nearest-neighbor square lattice. In contrast, our approach divides the network into clusters where qubits within each cluster---as well as those in adjacent clusters---exhibit all-to-all logical connectivity. Due to this significantly higher connectivity, our architecture can reduce the total circuit depth during logical-level compilation, potentially offsetting the serialization penalty when compared to the surface code.

{\bf Summary.} Our results show a clear trade-off between qubit overhead and time overhead as we vary the density of surface-code blocks. Relative to a pure surface-code architecture, our protocol achieves a smaller qubit overhead but incurs a larger time overhead. Regarding the enlarged time overhead, we make the following remarks:
\begin{itemize}
\item[i)] {\it Relevance of qubit overhead for early fault-tolerant devices.} For early-stage fault-tolerant hardware, physical qubit overhead represents the primary bottleneck. Such devices will have only a modest number of physical qubits yet must encode more than the minimum number of logical qubits required to reach meaningful quantum advantage. Consequently, qubit overhead is critical and may determine whether fault-tolerant quantum computation is feasible at all on near-term hardware, i.e. more important than time overhead.
\item[ii)] {\it Use of surface-code patches reflects state-of-the-art approaches for realistic architectures.} Introducing surface-code patches---and thereby incurring additional time overhead---is consistent with state-of-the-art techniques for practical architectures, especially for superconducting qubit platforms, such as the protocol in Ref.~\cite{yoder2025tour}.
\item[iii)] {\it Protocols with asymptotically optimal spacetime overhead.} There are theoretical schemes that achieve time overheads as low as $O(d^{1+o(1)}) = O(\log^{1+o(1)}(1/\epsilon))$ while maintaining constant qubit overhead~\cite{Nguyen2024,zhang20253}. However, such protocols typically assume idealized architectures with unrestricted qubit connectivity---assumptions not applicable to realistic two-dimensional hardware. In contrast, our protocol operates under physically motivated locality and connectivity constraints.
\end{itemize}

\section{Comparison with Existing Logical Measurement Protocols}

Our main aim is to develop a method for implementing logical operations under stringent constraints on the physical qubit network: specifically, only short-range couplings on an open-boundary two-dimensional qubit array with transversal symmetry, and allowing only measurements of the template stabilizers that generate the underlying planar code. These constraints are motivated by minimizing the hardware requirements and physical error per stabilizer-measurement cycle. Under these constraints on connectivity and allowed operations, our protocol is, to the best of our knowledge, the only known approach that enables lattice-surgery-type logical operations. Existing methods---including CKBB scheme~\cite{Cohen2022}, gauging measurement~\cite{Ide2024,Williamson2024}, and devised sticking~\cite{Zhang2025}---are not applicable under these restrictions. Therefore, the primary advantage is that our protocol yields robust logical measurements without requiring long-range interactions or additional hardware complexity.

For example, directional codes can be realized on two-dimensional square lattices with strictly nearest-neighbor connectivity---the same physical network architecture used for surface codes~\cite{geher2025directional}. Despite utilizing identical hardware, directional codes offer significantly higher encoding rates. By employing our protocol, we can design fault-tolerant logical measurement circuits for directional codes while strictly adhering to nearest-neighbor interaction constraints (see Fig.~\ref{fig:NANO}).

To implement directional-code logical measurements on nearest-neighbor qubit arrays, the deformed codes produced by our code craft framework are realized by deactivating (cutting) specific qubits while maintaining the global regularity of the iSWAP-gate pattern. This deactivation is achieved by initializing the deactivated qubits into appropriate basis states. Given that an iSWAP gate can be decomposed into SWAP and controlled-NOT components (up to single-qubit Clifford gates), a qubit can be neutralized by rendering its associated controlled-NOT components trivial. Specifically, if a qubit serves as the control for a gate, it is prepared in the $\ket{0}$ state; if it serves as the target, it is prepared in the $\ket{+}$ state.

Numerical investigations across various codes (with distances up to $d=21$) illustrates that our protocol requires an ancilla system size comparable to that of the original code block. Specifically, for an $[[n,k,d]]$ code, the original block comprises approximately $2n$ physical qubits ($n$ data qubits and about $n$ ancilla qubits for syndrome extraction). Our fault-tolerant logical measurement protocol requires an additional $2\chi n$ physical qubits---the ancilla system---where $\chi$ is a factor ranging between $0.155$ and $1.582$. Consequently, the total cost is approximately $2(1+\chi)n$ physical qubits.

For comparison, the CKBB protocol requires an ancilla system of approximately $4wd$ physical qubits, where $w \geq d$ is the weight of the measured logical operator. This results in a total qubit count of $2n + 4wd$, where $4wd$ is typically comparable to $2n$ even for $w = d$. Gauging measurement protocol offers better asymptotic scaling of $O(w\mathrm{polylog}(w))$, leading to a total qubit requirement of $2n + O(w\mathrm{polylog}(w))$. The devised sticking protocol, while capable of addressing multiple operators, requires a total qubit count significantly exceeding $2n$.

Therefore, even in an idealized scenario where existing protocols utilize negligible ancilla resources, our approach introduces an overhead factor of $1+\chi$ (ranging from $1.155$ to $2.582$ based on our numerical results). We contend that this modest increase in qubit overhead is a justified trade-off for the significantly improved compatibility with planar physical architectures.

Regarding circuit depth and fault tolerance, established proofs in Refs.~\cite{Zhang20251,Cross2024,Williamson2024} show that a logical measurement is fault-tolerant with a circuit-level distance of $d$ provided that it includes $d$ rounds of syndrome extraction on the deformed code, and that the deformed code itself possesses a distance of $d$. This requirement is a universal feature shared by the CKBB protocol, gauging measurement, devised sticking, and our planar protocol. Consequently, all these protocols exhibit the same $O(d)$ time overhead and provide equivalent fault-tolerant guarantees.

Regarding the applicability, guided by the topological framework of our method, the \textit{stretching--cutting--painting} procedure is broadly applicable to a wide range of planar topological codes, including the family of planar BB codes. This is motivated by the universal encoding scheme for such codes: a code block consists of a square manifold with two $m$-condensed and two $e$-condensed boundaries, analogous to the surface code ~\cite{Liang2025, liang2025generalized,Fowler2012}. Within this framework, an $X$ logical operator corresponds to a string operator connecting the $m$-condensed boundaries. By extending the lattice and modifying the boundary stabilizers---steps we refer to as stretching and boundary cutting---we can convert an $e$-condensed boundary into an $m$-condensed one, merging the two pre-existing $m$-condensed boundaries into a single contiguous $m$-condensed boundary. This transformation results in a deformed code that incorporates the original $X$ logical operators into its stabilizer group, which is exactly the purpose of stretching and boundary cutting. Note that bulk cutting and painting are applicable to general codes (see the Discussion section in the main text). Therefore, our protocol is applicable to a wide range of planar topological codes.

\section{Comparison of logical operator painting with the final-measurement approach}

Remark 33 of Ref.~\cite{Williamson2024} proposes an alternative strategy for managing gauge degrees of freedom: inferring gauge-operator eigenvalues from the final ancilla-system measurements and utilizing them as auxiliary stabilizers for error detection. However, this approach lacks a formal guarantee of fault tolerance, even in the large-distance limit. In contrast, our logical operator painting framework rigorously ensures fault-tolerant operation, provided that the resulting code distance is sufficiently large.

Throughout this section, we focus on the measurement of $X$-type logical operators; the corresponding circuit is shown in Fig.~3 of the main text. In this case, logical operator painting and the final-measurement approach are utilized to improve error correction against $X$-type errors, thereby providing enhanced protection for the $Z$-logical operators.

\textbf{Distinction.} The two approaches yield distinct codes, resulting in different implementations.

Let the initial subsystem code be denoted by $\mathcal{C}_{g1} = (S_X, S_Z, L_X, L_Z, G_X, G_Z)$, where $S_{X/Z}$, $L_{X/Z}$, and $G_{X/Z}$ represent the generating sets of $X$- and $Z$-type stabilizers, logical operators, and gauge operators, respectively. We distinguish between the two approaches as follows:
\begin{itemize}
    \item Remark 33 of Ref.~\cite{Williamson2024}: By promoting the $Z$-type gauge operators to the stabilizer group, the initial subsystem code is transformed into a stabilizer code $\mathcal{C}_{s1} = (S_X, S_Z \cup G_Z, L_X, L_Z)$.
    \item Logical Operator Painting: We maintain the code as a subsystem code $\mathcal{C}_{g2} = (S_X, S_Z, L_X, L_Z', G_X', G_Z)$, but redefine the $Z$ logical operators as $L_Z' = \{l \cdot g(l) \mid l \in L_Z\}$, where $g: L_Z \to \langle G_Z \rangle$ is a mapping optimized to maximize the code distance; the $X$-gauge generating set is updated accordingly to preserve the necessary commutation relations.
\end{itemize}
While one could combine the two to form a stabilizer code $\mathcal{C}_{s2} = (S_X, S_Z \cup G_Z, L_X, L_Z')$, it is mathematically identical to $\mathcal{C}_{s1}$ because $g(l)$ are contained within the stabilizer group of the stabilizer codes.

The four codes---$\mathcal{C}_{g1}$, $\mathcal{C}_{s1}$, $\mathcal{C}_{g2}$, and $\mathcal{C}_{s2}$---each possess an associated code distance, denoted by $d_{g1}$, $d_{s1}$, $d_{g2}$, and $d_{s2}$, respectively. Since $\mathcal{C}_{s1}$ and $\mathcal{C}_{s2}$ are equivalent, $d_{s1} = d_{s2}$. The relationships between these distances are governed by three factors: $d_{s1} \geq d_{g1}$ due to the larger stabilizer group in $\mathcal{C}_{s1}$; $d_{g2} \geq d_{g1}$ because the painting procedure optimizes the code distance; and $d_{s2} \geq d_{g2}$ because $\mathcal{C}_{s2}$ has a larger stabilizer group. Consequently, we establish the distance hierarchy: $d_{s2} = d_{s1} \geq d_{g2} \geq d_{g1}$.

The distinction between $C_{s1}$ and $C_{g2}$ arises from two fundamental differences:
\begin{enumerate}
    \item Parity-check measurements: To implement $C_{s1}$, both $S_Z$ and $G_Z$ must be measured. If only $S_Z$ is measured, the distance $d_{s1}$ is not realized, and logical errors with weights between $d_{g1}$ and $d_{s1}$ remain undetectable. In contrast, implementing $C_{g2}$ requires only $S_Z$ measurements, yet all undetectable logical errors are guaranteed to have weights of at least $d_{g2}$.
    \item Decoding procedures: The decoding logic differs for the two codes. For $C_{s1}$, error decoding utilizes the measurement outcomes of both $S_Z$ and $G_Z$. For $C_{g2}$, decoding relies exclusively on the outcomes of $S_Z$.
\end{enumerate}

We now explicitly illustrate the operational distinction between these two approaches. Consider the following assumptions:
\begin{itemize}
    \item Let $s \in S_Z$ be a stabilizer, $l \in L_Z$ a logical operator, and $g \in G_Z$ a gauge operator.
    \item Suppose $s$ and $g$ overlap at a single qubit $q_1 \in \text{supp}(s) \cap \text{supp}(g)$, and let $q_2 \in \text{supp}(s)$ be another qubit in the support of the stabilizer ($q_2 \neq q_1$). 
    \item Assume $l$ is disjoint from both $s$ and $g$, i.e.,~$\text{supp}(l) \cap \text{supp}(s) = \text{supp}(l) \cap \text{supp}(g) = \emptyset$.
    \item Following the painting procedure, the redefined logical operator is $l' = l \cdot g \in L_Z'$.
    \item Suppose the final $Z$ measurements return eigenvalues of $-1$ for both $s$ and $g$.
\end{itemize}

For the stabilizer code $\mathcal{C}_{s1}$: Because both $s$ and $g$ are treated as stabilizers, the decoder processes the joint syndrome $(-1, -1)$. Given the overlap at $q_1$, the decoder may infer an error on $q_1$. The subsequent recovery operation on $q_1$ restores both $s$ and $g$ to the $+1$ eigenspace. Since $q_1 \notin \text{supp}(l)$, this correction does not affect the logical state of $l$.

For the subsystem code $\mathcal{C}_{g2}$: Only $s$ is treated as an active stabilizer. Upon observing the syndrome $s = -1$, the decoder may infer an error on $q_2$. The recovery operation on $q_2$ restores $s$ to $+1$. However, the eigenvalue of the gauge operator $g$ is still $-1$; accordingly, we apply a feedback gate to flip $l$.

This comparison demonstrates that the two approaches utilize distinct sets of syndrome information, leading to different recovery operations and feedback mechanisms.

Finally, taking the operators shown in Fig.~4 in the main text as examples, if the gauge operator $Z_{III}$ detects an error, feedback is applied to the logical operator $Z_{I}$. However, as we have explained, this mechanism is fundamentally distinct from a standard stabilizer code picture where $Z_{III}$ is taken as a stabilizer. It is our algebraic painting procedure that identifies the correlation between error events on $Z_{I}$ and $Z_{III}$, a relationship that is not revealed by the error decoding according to the stabilizer code $\mathcal{C}_{s1}$.

In $\mathcal{C}_{g1}$, $Z_{I}$ is the defined logical operator, and $X_C$ is indeed an undetectable logical error that determines the code distance. However, in our painted code $\mathcal{C}_{g2}$, the logical operator is redefined as $Z_{II}$. Because $X_C$ commutes with $Z_{II}$, it acts as an undetectable trivial error (assume it commutes with all other $Z$-type logical operators). Consequently, $X_C$ no longer affects the code distance.

\textbf{Fault tolerance.} The code $\mathcal{C}_{g2}$ is characterized by two essential properties:
\begin{itemize}
    \item When decoded using only $S_Z$ measurement outcomes, all logical errors have weights of at least $d_{g2}$.
    \item The stabilizers $S_Z$ have low weight, and the code is a qLDPC code.
\end{itemize}
As a consequence of these properties, the error correction procedure is fault-tolerant under local stochastic noise~\cite{Gottesman2014}.

We now compare the approach in Remark 33 of Ref.~\cite{Williamson2024} with our logical operator painting framework in terms of fault tolerance. Note that our purpose is to implement a fault-tolerant logical measurement. Suppose we are given a set of candidate deformed codes $\{\mathcal{C}_{g1}\}$ intended to facilitate a specific logical measurement.
\begin{itemize}
\item Remark 33 of Ref.~\cite{Williamson2024}: This methodology generates a set of stabilizer codes $\{\mathcal{C}_{s1}\}$ with corresponding code distances $\{d_{s1}\}$. While one might select a code from this set based on a large distance $d_{s1}$, this distance is only realized if the promoted gauge operators are treated as active stabilizers throughout entire the circuit. If these operators are only measured at the final stage, the distance $d_{s1}$ does not necessarily guarantee fault tolerance.
\item Logical Operator Painting: In contrast, our approach generates a set of gauge codes $\{\mathcal{C}_{g2}\}$ with distances $\{d_{g2}\}$. Here, the code distance is evaluated specifically in the context of the available stabilizers. As we have demonstrated, selecting a candidate code with a sufficiently large $d_{g2}$ guarantees fault tolerance. This is because the distance $d_{g2}$ represents a true lower bound on the weight of undetectable logical errors when decoding relies exclusively on the low-weight $S_Z$ measurements.
\end{itemize}

Finally, we conjecture that the synergy between these two approaches may yield optimal performance. Specifically, one could select the candidate code according to the distance $d_{g2}$ instead of $d_{s1}$, while simultaneously incorporating the final-measurement syndrome information from the gauge operators $G_Z$ into the decoding process. Since this hybrid approach utilizes an augmented set of syndrome data, it is expected to provide slightly superior error suppression. However, as the realized performance remains sensitive to the choice of decoding algorithm and specific hardware constraints, we present this as a hypothesis.

As a concrete example, when constructing a deformed code via our code craft protocol, selecting an appropriate ancilla system size is critical. As illustrated in Table~\ref{table:162}, the resulting code distance is a function of the ancilla size; for each size, we report two distances, for $\mathcal{C}_{g1}$ and $\mathcal{C}_{g2}$, respectively. We note that these two codes result in almost identical logical-measurement circuits (shown in Fig.~3 of the main text). The only distinction lies in the feedback gate $V$, which depends on the choice of logical operators: the feedback gate restores the $Z$-logical information of the original code by accounting for the specific form of the deformed-code logical operators ($L_Z$ or $L_Z'$). Furthermore, if the final-measurement approach is adopted, a third feedback strategy arises, wherein the eigenvalues of the $Z$-gauge operators are utilized as additional syndrome information for error correction (corresponding to code $\mathcal{C}_{s1}$). The critical insight is that the ancilla size must be chosen based on the distance of $\mathcal{C}_{g2}$, rather than $\mathcal{C}_{s1}$, to strictly guarantee fault tolerance. Once this size---and consequently the deformed-code stabilizer group---is fixed, recovery strategies based on either $\mathcal{C}_{g2}$ or $\mathcal{C}_{s1}$ are both fault-tolerant.

% \bibliography{references.bib}

\end{widetext}

\end{document}